\documentclass[12pt]{JHEP3}
\usepackage{amssymb, amsmath, amsopn, amsthm}
\usepackage{epsfig}
\usepackage{graphics}
\newcommand{\be}{\begin{equation}}
\newcommand{\ee}{\end{equation}}
\newcommand{\ti}{\tilde}
\newcommand{\f}{\frac}
\newcommand{\p}{\partial}
\newcommand{\tir}{\tilde{r}}

\title{Bianchi Attractors: A Classification of Extremal Black Brane Geometries}
\author{Norihiro Iizuka$^{1}$, Shamit Kachru$^{2}$, Nilay Kundu$^{3}$, Prithvi Narayan$^{3}$, \newline
Nilanjan Sircar$^{3}$ and Sandip P. Trivedi$^{3}$

  ~\\
 
 $^1$Theory Division, CERN, CH-1211 Geneva 23, Switzerland \\
norihiro.iizuka@cern.ch \\
$^2$ Stanford Institute for Theoretical Physics, Department of Physics\\
and Theory Group, SLAC National Accelerator Laboratory\\
Stanford University, 
Palo Alto, CA 94305, USA\\
skachru@stanford.edu \\
$^3$Tata Institute for Fundamental Research \\
Homi Bhabha Road, Mumbai 400005, India\\
nilay.tifr@gmail.com, prithvi.narayan@gmail.com, nilanjan@theory.tifr.res.in,  trivedi.sp@gmail.com

\vspace{0.1cm}

}

\abstract{ 
Extremal black branes are of interest because they correspond to the ground states of field theories
at finite charge density in gauge/gravity duality.
 The  geometry of such a brane need not be translationally 
invariant in the spatial directions along which it extends. A less restrictive requirement is that 
of homogeneity, which still 
allows points along the spatial directions to be related to each other by symmetries. 
In this paper, we find large new classes of homogeneous but anisotropic extremal black
brane horizons, which could naturally arise in gauge/gravity dual pairs.
In $4+1$ dimensional spacetime, we show that such homogeneous black brane solutions are classified by the Bianchi 
classification, which is well known in the study of cosmology, and fall into nine classes.
In a system of Einstein gravity with negative cosmological term coupled to one or two massive Abelian gauge fields, we find 
solutions with an additional scaling symmetry, which could correspond to  
the near-horizon geometries of such extremal black branes. 
These solutions realize many of the Bianchi  classes.
In one case, we construct
the complete extremal solution which asymptotes to AdS space.

}

\preprint{CERN-PH-TH-2012-005 \\ TIFR/TH/12-02 \\SU-ITP-12/04}
\begin{document}
\tableofcontents
\section{Introduction}
Nature can  exist in many varied and beautiful phases. 
String theory  too  has many varied and beautiful  phases, which correspond to the huge landscape of vacua 
in the theory. It is natural to hope that some of the phases of string theory might 
help  describe  those in Nature. 
 This hope has  spurred recent progress   attempting, for example, 
  to build connections between string theory
 and condensed matter physics. 

Some of the work in this direction has involved the study of black branes in gravity theories,
with a particular focus on their role as holographic duals to field theories at finite temperature
and/or chemical
potential.
Extremal branes  are particularly interesting,
since  they  correspond to zero temperature ground states.  
Without a temperature 
  quantum fluctuations come into ``their own'', especially in 
 strongly correlated systems,  leading to interesting and novel phenomena like
quantum phase transitions. 
The   description of these ground states 
 in terms of an extremal  brane allows such  effects  to be studied in a non-trivial  and  often tractable 
mean field approximation.

So far only a  few different kinds of extremal brane solutions have been found. The analogy with the phases of 
matter, mentioned above,  would suggest that many more should exist. 
The main point of this paper is to show that this expectation is indeed true. 
We will argue below  that the known types of extremal black branes are only  the  ``tip of the iceberg'',
and that there is a much bigger zoo of solutions,  obtainable in theories of gravity with reasonable matter, 
  waiting to be discovered. 

Symmetries are a good place to start in classifying the solutions of general relativity and also in classifying 
the phases of matter.  
The known extremal solutions mostly  have the usual  translational symmetries along the spatial
 directions in which the 
brane extends.
  The solutions we consider here differ by having a generalised version of translational invariance. 
Any two points along the spatial  directions can still be connected by a symmetry transformation,
but  the generators of the symmetries now do not necessarily commute with each other. 
This generalised notion of translations is well known in general relativity. Space-times with such  symmetries are said
to be  homogeneous. 

To be more precise, we will study brane solutions of the form
\be
\label{metrica}
ds^2=dr^2 -g_{00}dt^2+  g_{ij}dx^i dx^j
\ee
where $i,j =  1, \cdots d-1$  are the spatial  directions along which the brane extends.
 We will find that often, for reasonable 
 matter Lagrangians,
 there are  brane   solutions 
 where the  $x^i$ coordinates span a homogeneous space  with isometries which  do not commute.
Most of our discussion will be for the 
 case $d=4$ where the brane extends  in $3$ space dimensions \footnote{Towards the end of the paper
we will briefly discuss examples where the time direction could also be involved in the generalised 
translations.}. 
The different kinds of generalised translational symmetries that can arise along three spatial directions are
 well known in general relativity. They lie in the  Bianchi classification and  fall 
into  $9$ different classes, e.g., see \cite{LL}, \cite{SR}. 
These classes therefore classify all homogeneous brane solutions of the type eq.(\ref{metrica}). 

%We will show that  solutions lying in many of these $9$ classes can be obtained by coupling gravity to one or 
%two massive Abelian gauge fields in the presence of a cosmological constant.  

Of particular interest in the study of extremal branes  is their near-horizon geometry. 
 The near-horizon geometry encodes
information about the low energy dynamics of the dual  field theory and often turns out to be 
 scale invariant. 
The near horizon geometry is also often an attractor, with differences from the attractor
geometry far away 
dying out as one approaches the horizon. 
This feature corresponds to the fact that much  of the UV data in the field theory is 
often irrelevant in the IR.  
  These properties mean that the 
near-horizon geometry is often easier to find analytically than the full solution.  The attractor nature also 
makes the near-horizon geometry more universal than the full extremal solution.
Thus one is in the happy situation that the IR  region,  which  being more universal is of greater
 interest anyway, 
is also the region one can obtain with relative ease.

In this paper we will focus mostly on the near-horizon region. We will consider solutions which are 
 scale invariant in this region, along with being homogeneous  along the spatial  directions 
\footnote{ The scaling symmetry
 will correspond to translations in the radial coordinate $r$.
Thus, including the scaling symmetry makes the full $d+1$ dimensional space-time geometry homogeneous.}. 
We will show that  solutions lying in many of the $9$ Bianchi  classes mentioned above
 can be obtained by coupling gravity to  relatively simple kinds of matter. In the examples we consider,
one or two massive Abelian gauge fields in the presence of a negative cosmological constant will suffice. 
In one case we will also find  interpolating solutions for the full extremal brane which interpolate between 
the scaling near-horizon region and asymptotic $AdS$ space. 

A simple example of the kind of generalised translation invariance we have in mind arises in condensed matter 
physics when a non-zero momentum mode condenses, leading to a vector order parameter  which  varies  in 
a helical manner in space with a pitch determined by the momentum of the condensed mode. 
For example, in spin systems this is known to happen when a spin wave of non-zero momentum condenses, resulting 
in the magnetisation order parameter varying in a helical pattern (see \cite{condmatt1}, \cite{condmatt2}
 for a discussion). 
In superconductors, it has been argued that such spatially modulated phases can arise due to the FFLO instability
 \cite{FF1}, \cite{FF2}. Similar behavior can also occur in 
QCD \cite{Rubakov} and other systems \cite{Brazov}. 
 It has also been discussed recently 
in the context of AdS/CFT in \cite{DH}, \cite{ooguri} and \cite{gauntlettspace}. In the context of our discussion such situations lie in 
the Type VII class of the Bianchi classification. 
We will   discuss this case below  quite extensively  since it is  relatively simple and 
illustrates many of  the  features
which arise in the other classes as well. 

The paper is structured as follows. We begin with a discussion of the generalised translations and various 
 Bianchi classes in \S2. Then we discuss one illustrative example of  a  Bianchi Type VII near-horizon geometry 
in \S3. 
Other examples giving rise to Bianchi Types II, III, V, VI, and IX are discussed in \S4. 
A concrete example where a Type VII near-horizon geometry can arise from an asymptotically 
$AdS$ spacetime is given in \S5. A brief discussion of subtleties which might arise when time
is involved in the generalised translations is contained in \S6.
The paper ends  with some discussion in \S7. 
Important supplementary material is contained  in the Appendices \S\ref{AppA}-\S\ref{AppD}. 

Before closing the introduction, let us comment on related literature. 
There is a formidable body of work on brane solutions in the string theory and general relativity literature; for a recent review with further references,
see \cite{Marolf}.  
A classification of extremal black holes (as opposed to black branes), quite different
from ours, has been discussed in \cite{Reall}.
Our solutions can be viewed as black branes with new kinds of ``hair'';
simple discussions of how branes in $AdS$ space can violate the black hole no-hair
theorems are given in the papers on holographic superconductivity,
see e.g. \cite{gubser} and  \cite{horowitz1}  for
discussions with additional references.   Early examples of black branes with interesting
horizon structure were discussed in studies of the Gregory-Laflamme instability; the original
papers are 
\cite{Gregory} and a recent discussion appears in \cite{Pretorius}.  As we mentioned
previously, solutions with instabilities of the Type VII kind have   appeared already in the context of 
AdS/CFT duality
in the interesting papers \cite{DH, ooguri, gauntlettspace}.
Lifshitz symmetry has characterized one new type of horizon to emerge in holographic duals of field theories at finite charge density in many recent studies; this was discussed in \cite{KLM} and \cite{MT} (see also 
 \cite{narayan} for string theory and supergravity embeddings of such solutions).  
The attractor mechanism has also inspired a considerable literature. The seminal paper is \cite{FKS}.
For a recent review  with a good collection of references, see \cite{lectures}. Some references on attractors without
supersymmetry are \cite{Ga, Gibbons,Sen,attractors,Renata}. A  recent discussion 
of how the attractor mechanism may be related to new kinds of horizons for black branes, 
including e.g. Lifshitz solutions, appears in  \cite{KKS}.

\section{Generalised Translations and The Bianchi Classification}\label{simpleexample}
It is worth beginning with a simple example illustrating  
the kind of generalised translational symmetry we would like to explore. 
Suppose we are in three dimensional space described by coordinates $x^1,x^2,x^3$ and suppose the system of interest 
has the usual translational symmetries along the $x^2,x^3$ directions. It  also has 
 an additional symmetry  but  
 instead of being the usual  translation in the $x^1$ direction  it  is  now  
   a  translation  accompanied by a rotation in the $x^2-x^3$ plane.
The translations along the $x^2,x^3$ direction are generated by the  vectors fields, 
\be
\label{transa}
\xi_1=\partial_2, \ \  \xi_2=\partial_3,
\ee 
whereas the third symmetry is generated by  the vector field, 
\be
\label{transb}
\xi_3=\partial_1+x^2\partial_3-x^3\partial_2.
\ee
It is easy to see that the three transformations generated by these vectors  transform any point in the 
three dimensional plane to any other point in its immediate neighborhood. This property, which is akin to that 
of usual translations, is called homogeneity. However, the symmetry group we are dealing with here is  clearly 
different from usual translations, since  the commutators of the generators, eq.(\ref{transa},\ref{transb})  take the form
\be
\label{comalg}
[\xi_1,\xi_2]= 0;  \ \  [\xi_1,\xi_3]=\xi_2;  \ \   [\xi_2,\xi_3]=-\xi_1,
\ee
and do not vanish, as they would have for the usual translations \footnote{In fact this symmetry group is simply the 
group of symmetries of the two dimensional Euclidean plane, its two translations and one rotation.}. 

In this paper we will be interested in exploring  such  situations in more generality
  where the symmetries  are different from the usual translations but still preserve  homogeneity, allowing any  point
 in the system of interest to be transformed to any other by a symmetry 
transformation \footnote{In a connected space this follows from the requirement that  any point is
 transformed to any other point 
in its immediate neighborhood by a symmetry transformation.}.

When can one expect such a symmetry group to arise? 
Let us suppose that there is a scalar order parameter, $\phi(x)$, which specifies the system of interest. 
Then this order parameter must be invariant under the unbroken symmetries.  
The change of this order parameter under an infinitesimal transformation generated by the vector field
 $\xi_i$ is given by 
\be
\label{gtop}
\delta \phi =\epsilon \xi_i(\phi)
\ee
and this would have to vanish for the transformation to be a symmetry. It is easy to see that 
requiring that this is true for the 
three symmetry generators  mentioned above, eq.(\ref{transa}), eq.(\ref{transb}), leads to the condition 
 that $\phi$ is a constant independent of the coordinates, $x^1,x^2,x^3$. 
Now, such   a constant configuration for $\phi$ is invariant under the usual translation along $x^1$, generated by $\partial_1$,
besides being invariant also under a rotation in the $x^2-x^3$ plane (and  the other rotations). 
Thus we see that    with a scalar order parameter   a situation where the generalised translations  are unbroken 
 might as well be thought of as   one that  preserves 
the usual three translations along with additional rotations. 

Now suppose that instead of being a scalar  the order parameter specifying the system  is a vector $V$, 
which we denote by $V^i\partial_i$. Under the transformation generated by a vector field $\xi_i$ this transforms as 
\be
\label{vectrans}
\delta V=\epsilon [\xi_i,V].
\ee
Requiring that  these commutators vanish for all three $\xi_i$'s leads to the conditions, 
\be
\label{condva}
V^1={\rm constant}, \ \ V^2=V_0 \cos(x^1+\delta), \\ V^3=V_0\sin(x^1+\delta) 
\ee
where $V_0,\delta$ are constants independent of all coordinates. 
In other words the $V^2,V^3$ components are not constant  but rotate in the $2-3$ plane as one advances along the $x^1$
direction. 
In contrast, for the usual translations all three components $V^1,V^2,V^3$ would be constant and the behavior of the order parameter would be different. 
Thus we see that with a vector order parameter, instead of a scalar,
 one can have  situations where the generalised translational 
 symmetries generated by eq.(\ref{transa}), eq.(\ref{transb}) are preserved but the usual translation group is not. 
Such a situation could also arise if the order parameter were a tensor field. 

In fact  situations   with this type of  a vector order parameter are well known to arise in   
condensed matter physics as was mentioned in the introduction. For example this can happen in spin systems,
where the order parameter is the magnetisation.  
In the presence of parity violation in such a system  a helical spin density wave of appropriate wavelength 
 can be set up  \cite{condmatt1}, \cite{condmatt2}.

\subsection{The Bianchi Classification} \label{bianchi}
 
The example discussed above shares many features in common with  other   symmetry groups which preserve homogeneity.
  Luckily all such groups  in three dimensions have been classified and are easily found in the physics literature
 since they have been of interest in the study of homogeneous situations in cosmology \cite{LL}, \cite{SR}. 
It is known that there are  $9$  such inequivalent groups and they  are described by the Bianchi classification. 
The idea behind this classification is quite simple. As we have seen above the Killing vectors which generate the 
symmetries form a Lie Algebra. Homogeneity requires that there should be $3$ such Killing vectors in $3$ dimensions. 
The $9$ groups, and related $9$ classes in the Bianchi classification,  simply correspond to the $9$ inequivalent real Lie Algebras one can get with $3$ generators.   

The familiar case of translational invariance, which includes $AdS$ space and    Lifshitz spacetimes,
 corresponds to the case where the three generators commute and is called   Bianchi Class I.
The remaining  $8$ classes are different from this one and give rise  to new kinds of generalised translational 
symmetries. 
The symmetry group discussed above in eq.(\ref{transa}), eq.(\ref{transb}) falls within Class VII in the Bianchi classification, with the parameter $h=0$ in the notation of \cite{SR}, see page 112. We will refer to this case as Type $VII_0$ below. 
As mentioned above it corresponds to the group of symmetries of the two-dimensional Euclidean plane. 
Like the example above, in the more general case as well,  a scalar order parameter will not lead to a 
situation where the usual translations are broken. Instead a vector or more generally a  tensor order parameter is 
needed. Such  an order parameter, when it takes a suitable form, can preserve the symmetry group characterizing any of 
 the other $8$ classes. 

In Appendix A we give some details of the algebras for all $9$ Bianchi classes.
In our discussion below we will consider  specific  examples which lie in the Bianchi Classes, II, III, V, VI, VII and IX. 
Some extra details on these classes, including the  generators, invariant one-forms etc, can also 
be found in Appendix A.

\section{General Set-Up and A Simple Example}  \label{fivedim}
Here we turn to describing more explicitly  a spacetime metric incorporating the generalised translational symmetries 
described above 
and to asking when such a metric can arise as the solution to Einstein's equations of gravity coupled with matter. 
Mostly we will work in $5$ dimensions, this corresponds to setting
 $d=4$ in eq.(\ref{metrica}).
In addition,  to keep the discussion  simple, 
 we assume that the usual  translational symmetry along the time direction is preserved so that the metric is time 
independent  and  also assume that 
there are no off-diagonal components between $t$ and the other directions in the metric. This leads to 
\be
\label{metrcb}
ds^2=dr^2-a(r)^2 dt^2 + g_{ij} dx^i dx^j
\ee
with the indices $i,j$  taking values  $1, 2, 3$.
For any fixed value of $r,t$, we get a three dimensional subspace spanned by $x^i$.  We will take this 
subspace to  be a  
 homogeneous space, corresponding to one of the $9$ types in the Bianchi classification.

As discussed in \cite{SR}, for each of the $9$ cases  there are three invariant one forms, $\omega^i, i=1,2,3$,
 which are invariant under all $3$   isometries.  A metric expressed in terms of these one-forms with 
$x^i$ independent coefficients 
will be automatically invariant under the isometries.  For future reference we also note that these one-forms satisfy the relations
\be
\label{relonf}
d\omega^i={1\over 2} C^i_{jk}\omega^j\wedge \omega^k
\ee
where $C^i_{jk}$ are the structure constants of the group of isometries \cite{SR}. 

We take the  metric to  have one   additional isometry  which corresponds to scale invariance.
 An infinitesimal isometry of this type  will  shift the radial coordinate by $r\rightarrow r+ \epsilon$.
In addition we will take it to rescale the time direction with weight $\beta_t$, $t\rightarrow t e^{-\beta_t \epsilon}$,
and assume that it acts on the spatial coordinates $x^i$ such that  the invariant one forms, $\omega^i$ 
transform with weights $\beta_i$ under it, \footnote{The dilatation generator could act more generally than this. 
 A discussion of the  more general case  is left for the future.}
$\omega^i \rightarrow e^{-\beta_i \epsilon} \omega^i$.

These properties fix the metric to be of the form 
\be
\label{metricc}
ds^2 = R^2[ dr^2 - e^{2 \beta_t r} dt^2 + \eta_{ij} e^{(\beta_i +\beta_j)r}\omega^i\otimes \omega^j]
\ee
with  $\eta_{ij}$ being a constant matrix which is independent of all coordinates. 

In the previous section we saw that  a situation where the usual translations were not preserved but the 
 generalised translations were unbroken requires a vector (or possibly tensor) order parameter. A simple setting 
in gravity, which could lead to such a situation, is to consider an Abelian gauge field coupled to gravity.
We will allow the gauge field to have a mass in the discussion below.  
Such a theory with  a massive Abelian gauge field  is already known to give rise to  Lifshitz spacetimes 
\cite{KLM}, \cite{MT}.  Here we will find that it has in fact an even  richer set of 
solutions, exhibiting many possible phases and patterns of symmetries, as  parameters are varied. 

The action of the system  we consider is therefore  
\begin{equation}\label{action}
 S = \int d^5 x \sqrt{-g} \left\lbrace R + \Lambda - {1 \over 4} F^2  - {1 \over 4} m^2  A^2 \right\rbrace
\end{equation}
and has two parameters, $m^2, \Lambda$. We note that in our conventions $\Lambda>0$ corresponds to 
$AdS$ space. 

The three  isometries  along the $x^i$ directions, time translation and scale invariance 
restrict the form for the gauge potential to be\footnote{A possible $A_r \, dr$ term, where $A_r$ only depends on 
$r$ to be consistent with the symmetries,   can always be gauged away.}
\be
\label{fgp}
A=A_t e^{\beta_tr}dt + A_i e^{\beta_i r}  \omega^i
\ee
where $A_t, A_i$ are constants independent of all coordinates.

\subsection{A Simple Example Based on   Type $VII$}\label{simpletype7}
To construct a simple and explicit example we now   return to the  case
of Type $VII_0$ with which we began our discussion of generalised translational symmetry in \S2. We will show 
here that  a metric with this  type of  symmetry can arise  as a solution to Einstein's equations. 

The one-forms which are invariant under the isometries for this case  are given,   see \cite{SR}, page 112, by 
\begin{equation}\label{defineomega}
 \omega^1= dx^1 \ ; \ \omega^2=\cos(x^1) dx^2 + \sin(x^1) dx^3 \ ; \ \omega^3=- \sin(x^1)dx^2
+ \cos(x^1)dx^3.
\end{equation}
As discussed above we are assuming that the $x^i$ coordinates transform under the  scaling isometry  in such a way that 
the one-forms, $\omega^i$,  transform with definite weights. It is clear from eq.(\ref{defineomega}) that the only 
way this can happen is if $x^1$ is  invariant, and $(x^2,x^3) \rightarrow e^{-\beta \epsilon}(x^2,x^3)$,
 under $r\rightarrow r+\epsilon$,
so that $\omega^2, \omega^3$ have equal weights and transform as 
$(\omega^2,\omega^3)\rightarrow e^{-\beta \epsilon} (\omega^2,\omega^3)$.
To keep the discussion simple we also assume that the metric coefficients $\eta_{ij}$ which appear in
 eq.(\ref{metricc}) are diagonal. After some rescaling this allows the metric to be written as   
\be
\label{metricd}
ds^2=R^2[dr^2 -e^{2\beta_t r} dt^2+ (dx^1)^2 + e^{2\beta r} ((\omega^2)^2 + \lambda^2 (\omega^3)^2)].
\ee
The only unknowns that remain in the metric are the two constants,  $R, \lambda$ and the
two scaling weights, $\beta_t$ and $\beta$.

Before proceeding let us comment  on the physical significance of the parameter $R$. 
From eq.(\ref{defineomega}) we see that the coordinate distance in the $x^1$ direction needed to complete one 
full rotation in the $x^2-x^3$ plane is $2\pi$. From eq.(\ref{metricd}) it follows that the physical distance 
along the $x^1$ direction that is needed is $2\pi R$. Thus $R$ determines the pitch of the helix.

We will see below that a metric of the type in eq.(\ref{metricd})
 arises as a solution for the system, eq.(\ref{action}), when the gauge field takes 
the form
\be
\label{fgp2}
A=e^{\beta r}\left(\sqrt{\tilde{A_2}} \ \omega^2+\sqrt{\tilde{A_3}} \  \omega^3\right).
\ee

Let us begin with the gauge field equation of motion
\be
\label{gfeom}
d*_5F=-{1\over 2} m^2 *_5 A.
\ee
As discussed in Appendix B   this gives rise to the conditions eq.(\ref{condoneb}), eq.(\ref{condtwob}) (with the index $i$ taking values $2,3$).
For the metric eq.(\ref{metricd}),   eq.(\ref{condoneb}) is met.   
From eq.(\ref{defcab}),  (\ref{evcab}), (\ref{evcijk})
in Appendix B, and also from \cite{SR}, page 112, we have 
$k^1=0, k^2=k^3=-1$. 
Also comparing eq.(\ref{metricd}) and eq.(\ref{etaappb}) we see that
 $\lambda_1=\lambda_2=1, \lambda_3=\lambda$, and $\beta_1 = 0$,  $\beta_2 = \beta_3 = \beta$.  
If both $\tilde{A}_2$ and $\tilde{A}_3$ are non-vanishing
eq.(\ref{condtwob}) then gives
\begin{eqnarray}
\label{gfeom1}
\lambda^2 (m^2-2 \beta(\beta+\beta_t))+2 &=& 0\\
(m^2-2 \beta(\beta+\beta_t))+2\lambda^2  &=& 0.
\end{eqnarray}
Assuming $\lambda\ne 1$ these conditions cannot both be met. 
Thus we are lead to conclude that either $\tilde{A}_2$ or $\tilde{A}_3$ must vanish.  
We will set $\tilde{A}_3=0$ without loss of generality \footnote{From eq.(\ref{metricd}) and eq.(\ref{defineomega})
 we see that when $\lambda=1$  the symmetries in the $(x^1,x^2,x^3)$ directions are  enhanced to the 
usual translations and rotations.}, so that the gauge field is given by 
\be
\label{gff}
A=\sqrt{\tilde{A}_2}e^{\beta r} \omega^2
\ee
and the gauge field EOM gives  
\begin{equation}
\label{eomb}
 \lambda^2 (m^2-2 \beta(\beta+\beta_t))+2 = 0.
\end{equation}

Note that the solution we are seeking has five parameters, $R, \beta_t,\beta, \lambda,$ which enter in the metric and 
$\tilde{A}_2$, which determines the gauge field. These are all constants. 
 
Next,  turn to the Einstein equations
\be
\label{eequations}
R^{\mu}_{\nu}-{1\over 2} \delta^\mu_\nu R= T^\mu_\nu.
\ee
with
\be
T^\mu_\nu={1\over2}F^{\mu}_{\ \lambda} F_\nu^{\ \lambda}+{1\over4}m^2 A^\mu A_\nu + {1\over2}\delta^\mu_\nu\left(
\Lambda-{1\over4}F^{\rho \sigma} F_{\rho \sigma}-{1\over4}m^2 A^\rho A_\rho\right)
\ee
Along the $tt$, $rr$, and $x^1x^1$  directions these give:
\begin{eqnarray}
\f{2(1+\ti A_2)}{\lambda^2}+2 \lambda^2+\ti A_2 (m^2+2 \beta^2)+24 \beta^2
-4(1+\Lambda) &=& 0\\
 \f{2(1+\ti A_2)}{\lambda^2}+2 \lambda^2+\ti A_2 (m^2-2 \beta^2)+8 \beta
(\beta+2\beta_t)-4(1+\Lambda) &=& 0\\
\f{2(1+\ti A_2)}{\lambda^2}+2 \lambda^2-\ti A_2 (m^2+2 \beta^2)-8(3\beta^2+2
\beta_t \beta+\beta_t^2) -4(1-\Lambda) &=& 0.
\end{eqnarray}
The components  along the $x^2,x^3$ directions lead to 
\begin{eqnarray}
\f{2(3+\ti A_2)}{\lambda^2}-2 \lambda^2-\ti A_2 (m^2+2
\beta^2)+8(\beta^2+\beta_t \beta+\beta_t^2) -4(1+\Lambda) &=& 0\\
\f{2(1+\ti A_2)}{\lambda^2}-6\lambda^2-\ti A_2 (m^2+2 \beta^2)-8(\beta^2+\beta_t
\beta+\beta_t^2) +4(1+\Lambda) &=& 0.
\end{eqnarray}
In obtaining these equations we have set $R$, which appears as the overall scale in front of the metric
eq.(\ref{metricd}), to be unity. The equations can then be thought of as determining the dimensionful parameters
 $\Lambda, m^2, \tilde{A}_2$ in terms of $R$. 
Counting eq.(\ref{eomb}) these are $6$ equations in all.  One can check that only $5$ of these are independent. 
These $5$ equations determine the $5$ parameters $(\Lambda, \beta,\beta_t, \lambda, \tilde{A}_2)$, which then 
completely determines the solution.

Solving, we get:
\begin{eqnarray}
\label{valpara}
 m^2 &=& \f{2(11+2 \lambda^2-10\lambda^4+3\lambda^6)}{\lambda^2 (5
\lambda^2-11)} \\
\Lambda &=& \f{1}{50} \left(95
\lambda^2+\f{25}{\lambda^2}-\f{50}{\lambda^2-2}+\f{144}{5\lambda^2-11}-146\right)\\
\beta &=& \sqrt{2} \sqrt{\f{2-3\lambda^2+\lambda^4}{5\lambda^2-11}} \label{valparabeta}\\
\beta_t &=& \f{\lambda^2-3}{\sqrt{2}~ (\lambda^2-2)}
\sqrt{\f{2-3\lambda^2+\lambda^4}{5\lambda^2-11}} \label{valparabetat}\\
\ti A_2&=& \f{2}{2-\lambda^2}-2.
\end{eqnarray}
The first of these equations can be thought of as determining $\lambda$ in terms of $m^2$, the second then as determining $\Lambda$
in terms of $m^2$, and so on, all in units where $R=1$. 

Requiring  $\tilde{A}_2>0$ gives 
\begin{equation}
\label{rangeaa}
 1\le \lambda <\sqrt{2}.
\end{equation}
Actually eq.(\ref{valparabeta}) and eq.(\ref{valparabetat}) only determine $\beta$ and $\beta_t$  upto an overall
 sign. We have chosen the sign so that in  the range eq.(\ref{rangeaa})
 $\beta_t$ are positive, and   the $g_{tt}$ component of the metric in 
particular  vanishes at the horizon, $r\rightarrow -\infty$.
Also we note that in this range, $m^2$
 is a monotonic function
of $\lambda$, and takes values in the range $-2\le m^2 <1$, where
$m^2(\lambda=1)=-2$.

\subsubsection{Some Concluding Remarks} \label{conrem}
We end this section with some concluding remarks. 

An important question which remains   is about  the stability of the solution we have found
above. 
We leave this for future study except to note that 
 in general, negative values of $m^2$ do not  necessarily imply  an instability, 
as is well known from the study of $AdS$ space, as long as the magnitude of $m^2$ is not very large. 

It is also worth comparing our solution to the well known Lifshitz solution  obtained from the 
same starting point eq.(\ref{action}). 
This solution is discussed in Appendix C for completeness. 
We see that with a spatially oriented gauge field the Lifshitz solution 
 can arise only if $m^2$ is positive, eq.(\ref{lastlif}).
In the parametric range $1>m^2>0$ both the Type $VII_0$ solution and the Lifshitz solution are allowed. 

Finally, it is worth noting that like the Lifshitz case, \cite{horowitz2},  while curvature invariants are finite
in the solution found above the tidal forces experienced by a freely falling observer can diverge at  the horizon.
The curvature invariants for the solution take  values
\begin{eqnarray}
R^\mu_\mu & = & {123\over25}+{72\over275-125\lambda^2} +{1\over \lambda^2-2}-{5+39\lambda^4\over 10\lambda^2}\\
R^{\mu\nu} R_{\mu\nu} &  = & {(\lambda^2-1)^2\over4\lambda^4(22-21\lambda^2+5\lambda^4)^2}(
1452-1804\lambda^2+8229\lambda^4 \nonumber \\ &-&14304\lambda^6+10114\lambda^8-3204\lambda^{10}+381\lambda^{12})
 \\
R^{\mu\nu\sigma\rho}R_{\mu\nu\sigma\rho} & = & {(\lambda^2-1)^2\over4\lambda^4(22-21\lambda^2+5\lambda^4)^2}(
5324-6732\lambda^2+12063\lambda^4 \nonumber \\ &-&19128\lambda^6 +14126\lambda^8-4700\lambda^{10}+583\lambda^{12})
\end{eqnarray}
and are indeed finite\footnote{All three of the curvature scalar diverges at 
$\lambda=0, \ \pm\sqrt{2}, \ \pm\sqrt{{11\over5}}$, \ but 
these values lie outside the range of validity of the solution given by eq.(\ref{rangeaa}).}. 
To see that the tidal forces can diverge we note that a
 family of geodesics can be found for the metric eq.(\ref{metricd}) where $x^1, x^2, x^3$ take  any constant values and $(r,t)$
vary with a functional dependence on the proper time, $\tau$, given by
  $(r(\tau),t(\tau))$ that is independent of these constant values.
Now take any two geodesics with the same $(r(\tau),t(\tau))$ which are separated along $x^2,x^3$.
The proper distance separating them in the $x^2,x^3$ directions  vanishes rapidly as the
 geodesics go
 to the horizon, $r\rightarrow -\infty$ leading to a diverging tidal force. 

It is worth trying to find the small temperature  deformation of  the Type VII solution above.
Such a deformation will probably allow us to control the effects of these diverging tidal forces. 
 
%%%%%%%%%%%%%%%%%%%%%%%%%%%%%%%%%%%%%%%%%%%%%%%%%%%%%%%%%%%%%%%%%%%%%%%%%%%%%%%%
%%%%%

%%%%%%%%%%%%%%%%%%%%%%%%%%%%%%%%%%%%%%%%%%%%%%%%%%%%%%%%%%%%%%%%

\section{Solutions of Other Bianchi Types}

The  discussion so far  has mostly  focused on examples of Bianchi Type VII.
 Solutions which lie in Bianchi Type I, e.g. Lifshitz solutions, are of course well known \cite{KLM}, \cite{MT}.
In this section we will give examples of some of the other Bianchi types.
Our purpose in this section  is not to give a very thorough or
complete discussion of these other classes, but rather  to illustrate how easy it is to obtain solutions
of these other types as well.

As in the discussion in section \S\ref{fivedim} for Type VII we   focus on solutions which have in addition the usual
time translational invariance as well as the  scaling symmetry  involving a translation in the radial direction.
 The presence of the scaling symmetry
 will simplify the analysis.  Also, such scaling solutions are often    the attractor end points
of more complete extremal black brane geometries. Accordingly we expect 
 the scaling solutions we discuss to  be more general than the
 specific systems we consider, and  to arise in a wide variety of settings.

A remarkable feature is that so  many different types of  solutions can be obtained from  relatively
 simple gravity + matter Lagrangians. Mostly, we will consider the system  described in eq.(\ref{action}) consisting of only one
massive Abelian gauge field and gravity. Towards the end, to get a non-trivial Type IX case, we will consider
two massive gauge fields.

Our analysis has two important limitations. We will not investigate the stability of these solutions. 
And we will not investigate whether the scaling solutions discussed in this section
 can be obtained as IR end points of more general geometries which are say asymptotically $AdS$. We leave these questions for the future. 
For the  Type VII case discussed in the previous section,  we will construct an example in \S5
 of a full extremal solution which asymptotes to 
$AdS$ space. 

The discussion in this section is a bit brief since it shares many similarities with the Type VII case studied in section \S\ref{fivedim}.

\subsection{Type II}
We begin by considering an example which lies in Type II. The system we consider is given by eq.(\ref{action}).
The symmetry group for Type II  is the Heisenberg group. The  generators in a convenient basis are given in Appendix A.
The invariant one-forms are
\be
\omega^1=dx^2-x^1 dx^3, \hspace{5mm}\omega^2=dx^3,\hspace{5mm}\omega^3=dx^1.
\ee
We take the metric to be of the form eq.(\ref{metricc}) and diagonal, 
\begin{equation}
 ds^2=R^2[dr^2-e^{2 \beta_t r}dt^2+e^{2 (\beta_2+\beta_3) r} (\omega^1)^2
+e^{2\beta_2 r}(\omega^2)^2+e^{2\beta_3 r}(\omega^3)^2].
\end{equation}
We take the massive gauge field to  be along the time direction
\be
\label{simplegf}
 A=\sqrt{A_t} \  e^{\beta_t r} \ dt. 
\end{equation}

As in the discussion of section \S\ref{simpletype7} we will find it convenient to set $R=1$.
The gauge field equation of motion is
\begin{equation}
m^2-4(\beta_2+\beta_3) \beta_t=0.
\end{equation}
The rr, tt, 11, 22, 33 components of the trace reversed Einstein equations\footnote{
Trace reversed Einstein equations in $d + 1$ space-time dimensions are  
\be
\nonumber
R^{\mu}_{\nu}= T^\mu_\nu - \frac{1}{d-1} \delta^\mu_\nu \, T \,, \quad 
T = T_{\mu\nu} g^{\mu\nu} \,.
\ee
} are:
\begin{eqnarray}
6(\beta_2^2+\beta_2\beta_3+\beta_3^2)-(A_t-3)\beta_t^2-\Lambda=0\\
A_t(3m^2+4\beta_t^2)+4 (-3 \beta_t (2 (\beta_2 + \beta_3) + \beta_t) +
\Lambda)=0\\
3 + 12 \beta_3 (\beta_2 + \beta_3) + 6 \beta_3 \beta_t +   A_t \beta_t^2 - 2
\Lambda=0\\
12 (\beta_2 + \beta_3)^2 + 6 (\beta_2 + \beta_3) \beta_t +   A_t \beta_t^2 - 3 -
2 \Lambda=0\\
3 + 12 \beta_2 (\beta_2 + \beta_3) + 6 \beta_2 \beta_t +  A_t \beta_t^2 - 2
\Lambda=0.
\end{eqnarray}
We have 6 parameters: $m^2, \Lambda, \beta_t, \beta_2, \beta_3, \ \text{and}
\ A_t$.
There are 5 independent equations, so we will express the 5 other parameters in terms of
$\beta_t$.

The solution is given as,
\begin{eqnarray}
 m^2=\beta_t \left(-\beta_t + \sqrt{16 + \beta_t^2}\right), \hspace{10mm}
\Lambda= {1\over16} \left(72 + 19 \beta_t^2 - 3 \beta_t \sqrt{16 + \beta_t^2}\right), \\
A_t= {19\over8} - {3 \sqrt{16 + \beta_t^2}\over 8 \beta_t}, \hspace{10mm}
 \beta_2 = \beta_3={1\over8} \left(-\beta_t + \sqrt{16 + \beta_t^2}\right).
\end{eqnarray}
Note that $\ A_t \ge 0, \ \beta_t > 0$ implies
\begin{equation}
 \beta_t \ge {3\over \sqrt{22}}.
\end{equation}

\subsection{Type VI, III and V}\label{type356}
Next we consider the three classes Type VI, III and V.
Type VI is characterized by one parameter, $h$.
The Killing vectors take the form, see Appendix A,
\be
\label{kvf}
\xi_1=\partial_2, \, \xi_2=\partial_3, \, \xi_3=\partial_1+ x^2\partial_2+ h x^3\partial_3 \,.
\ee
When $h=0$ we get Type III and when $h=1$ we get Type V.
Thus these two classes can be though of as limiting  cases of Type VI.

Looking at the Killing vectors in eq.(\ref{kvf}) we see that translations along the $x^2,x^3$ directions are of the
 conventional kind, but a translation along $x^1$ is accompanied by a rescaling along both $x^2$ and $x^3$ with weights
unity and $h$ respectively.

The invariant one-forms are
\be
\omega^1=e^{-x^1}dx^2, \ \omega^2=e^{- h x^1}dx^3, \ \omega^3=dx^1.
\ee

We consider again the system of a massive Abelian gauge field, eq.(\ref{action}).
We take the metric to be
\begin{equation}\label{type6metric}
 ds^2= R^2[dr^2-e^{2 \beta_t r}dt^2+(\omega^3)^2+e^{2 \beta_1 r}(\omega^1)^2+e^{2 \beta_2 r}(\omega^2)^2]
\end{equation}
and the gauge field to be
\begin{equation}
 A=\sqrt{A_t} \  e^{\beta_t r} \ dt.
\end{equation}
We also set $R=1$ below.

The gauge field equation of motion is then
\begin{equation}
m^2-2(\beta_1+\beta_2) \beta_t=0.
\end{equation}
The rr, tt, 11, 22, 33 component of the trace reversed Einstein equations are
\begin{eqnarray}
 3 \beta_1^2 + 3 \beta_2^2 - (A_t-3) \beta_t^2 - \Lambda=0\\
\beta_t (\beta_1 + \beta_2 + \beta_t) -  {1\over12} (3 A_t m^2 + 4 A_t \beta_t^2
+ 4 \Lambda)=0\\
6 + 6 h + 6 \beta_1 (\beta_1 + \beta_2) + 6 \beta_1 \beta_t +  A_t \beta_t^2 - 2
\Lambda=0\\
6h + 6 h^2 + 6 \beta_2 (\beta_1 + \beta_2) + 6 \beta_2 \beta_t +  A_t \beta_t^2
- 2 \Lambda=0\\
6 + 6 h^2 + A_t \beta_t^2 - 2 \Lambda=0.
\end{eqnarray}
Whereas the r3 component of Einstein's equation gives
\begin{equation}
 \beta_1+h \beta_2=0.
\end{equation}

The resulting solution is  conveniently expressed as follows:
\begin{eqnarray}
 m^2 &=& 2(1-h)^2 (1-\beta_2^2) \label{onevii} \\
\Lambda &=& 4 h + {(1-h)^2 \over \beta_2^2} + 2 (1-h+h^2) \beta_2^2 \label{twovii}\\
A_t &=& { 2 (1-h)^2 - 4 (1-h+h^2) \beta_2^2\over (1-h)^2 (1 - \beta_2^2)} \label{atvii}\\
\beta_1 &=& -h \beta_2 \\
\beta_t &=&   {(1-h) (1 - \beta_2^2) \over \beta_2}.
\end{eqnarray}

Choosing the radial coordinate so that the horizon lies at $r\rightarrow -\infty$, we get
$\beta_t>0$. If we require that $\beta_1,\beta_2\ge0$ also, and impose that
$A_t\ge 0$, as is required for the solution to exist, we get
the constraints:
\begin{equation}
\label{consvii}
 h \le 0 \hspace{10mm} \text{and} \hspace{10mm} 0<  \beta_2 \le {1-h \over
\sqrt{2} \sqrt{1-h+h^2} }.
\end{equation}

Having chosen a value of $h$ which satisfies eq.(\ref{consvii})
and thus a Bianchi Type VI symmetry, we can then pick a value of $\beta_2$ also meeting eq.(\ref{consvii}).
The rest of the equations can then be thought of as follows. Eq.(\ref{twovii}) determines
$R$ in units of $\Lambda$, eq.(\ref{onevii}) then determines $m^2$ required to obtain
this value of $\beta_2$ and the remaining equations determine the other parameters that enter in the solution.

Let us briefly comment on the limits which give Type V and Type III next.
\subsubsection{Type V}\label{type5}
To obtain this class we need to take the limit  $h\rightarrow 1$. To obtain a well defined limit where
$A_t$ does not blow up requires, eq.(\ref{atvii}), that $\beta_2\rightarrow 0$
keeping $\beta_t/(1-h)$ fixed.

It is easy to see that the solution can then be expressed as follows:
$m^2\rightarrow 0$, $\beta_1=0$ and
\be
\label{fv}
\Lambda=4 + \beta_t^2 \hspace{5mm} \text{and} \hspace{5mm} A_t=2 \
{\beta_t^2-2\over \beta_t^2}
\ee
with the condition,
\begin{equation}
 \beta_t^2 \ge  2.
\end{equation}

After a change of coordinates $x^1=\log(\rho)$ the metric becomes,
\begin{equation}
 ds^2=\left[dr^2-e^{2 \beta_t r}dt^2\right]+\left[{d\rho^2+(dx^2)^2+(dx^3)^2 \over
\rho^2}\right].
\end{equation}
We see that this is simply $AdS_2\times EAdS_3$, where $EAdS_3$ denotes the three-dimensional space of constant negative curvature obtained from the  Euclidean continuation of  $AdS_3$.

The resulting limit is in fact a one parameter family of  solutions   determined by the charge density.
To see this note that one can think of the first equation in eq.(\ref{fv})
as determining $R$ in terms of $\Lambda$, and the second equation as determining $\beta_t$ in terms of
$A_t$ which fixes the charge density.
$R$ determines the radius of $EAdS_3$ and $\beta_t$ then determines the radius of $AdS_2$.

Also, in this example  the symmetries of the spatial manifold are enhanced from those of Type V to the full
$SO(3,1)$ symmetry group of  $EAdS_3$ \footnote{One natural way in which such a solution can be obtained
is as the near horizon geometry of an extremal black brane in asymptotic $AdS_5$ space with non-normalizable metric deformations turned on so that the boundary theory lives in $EAdS_3$ space. One expects that the near- horizon
geometry can arise without such  a metric non-normalizable deformation being turned on as well.}.

\subsubsection{Type III}
The limit $h\rightarrow 0$ is straightforward.
One gets
\begin{eqnarray}
m^2&=&2(1-\beta_2^2) \\
\Lambda &=& {1\over \beta_2^2}+2\beta_2^2 \\
A_t & = & {2-4\beta_2^2\over 1-\beta_2^2}\\
\beta_1 & =&0 \\
\beta_t & = & {1-\beta_2^2\over \beta_2}.
\end{eqnarray}
The metric is
\be
\label{metiii}
ds^2=dr^2-e^{2\beta_t r} dt^2 + e^{2\beta_2 r}(dx^3)^2 + {1\over \rho^2} (d\rho^2+(dx^2)^2)
\ee
where $x^1=\log(\rho)$. We see that $\rho, \ x^2$ span two dimensional $EAdS_2$ space.

\subsection{Type IX and Type VIII}

In the Type IX case the symmetry group is $SO(3)$. The invariant one-forms are,
\begin{eqnarray}
 \omega^1&=& -\sin(x^3)dx^1+\sin(x^1) \cos(x^3)dx^2\\
\omega^2&=& \cos(x^3)dx^1+\sin(x^1) \sin(x^3)dx^2\\
\omega^3&=&\cos(x^1)dx^2+dx^3.
\end{eqnarray}
%We will consider in the following section, $x^1=x, \ x^2=y, \ x^3=z$ for convenience.

We consider a metric  ansatz
\be
ds^2=dr^2-e^{2 \beta_t r}dt^2 +(\omega^1)^2+(\omega^2)^2+ \ \lambda \ (\omega^3)^2.
\ee
This corresponds to $AdS_2\times  \rm{Squashed} \  S^3$.
We take the system to have  two gauge fields. One will be massless with only its time component turned on, and the other
will have mass $m^2$ and an expectation value proportional to the one-form $\omega^3$: 
\be
A_1=\sqrt{A_t} e^{\beta_t r} dt, \hspace{5mm} A_2=\sqrt{A_s} \omega^3=\sqrt{A_s}(\cos(x^1)dx^2+dx^3).
\ee

The equation of motion for the first gauge field is automatically satisfied, while the equation
of motion for the second yields
\be
m^2+2\lambda=0.
\ee
The independent trace reversed Einstein equations are
\begin{eqnarray}
 A_s+2[(A_t-3)\beta_t^2+\Lambda]=0\\
6-2A_s-A_t\beta_t^2-3\lambda+2\Lambda=0\\
A_s-A_t\beta^2-{3\over 2\lambda} A_s m^2+3\lambda+2\Lambda=0.
\end{eqnarray}
%First one comes from rr and tt components, 2nd one from 11,22 component and third one is the 33 component
% of the Einstein equations.

Solving these four equations we express $m^2,A_t,A_s,\Lambda$ in terms of $\beta_t, \lambda$. The solution is
\begin{eqnarray}
 m^2=-2\lambda; \hspace{10mm} A_t={1+2\beta_t^2 \over \beta_t^2}; \\
A_s=1-\lambda; \hspace{10mm} \Lambda={1\over 2}(2\beta_t^2+\lambda-3).
\end{eqnarray}
Then, $A_s\ge0 \ ,\Lambda>0, \ \beta_t>0$ implies
\be
\lambda\le1; \hspace{15mm} \beta_t>\sqrt{{3-\lambda \over 2}}.
\ee

Finally, let us briefly also comment on the Type VIII case \footnote{In the Type IX case studied above
 the spatial directions span a squashed $S^3$ which  is compact. In the Type VIII case the corresponding
manifold is non-compact and is more likely to arise starting with  an asymptotic geometry where the
spatial directions are along $R^3$ with no non-normalizable
metric perturbations.}. Here the symmetry group is $SL(2,R)$.
An example would be the case already considered in sec \S\ref{type356} in which case the spatial directions span $EAdS_3$ and
the  group of symmetries of the spatial manifold is enhanced to $SO(3,1)$.
However, we expect that more interesting examples, where the symmetry is only  $SL(2,R)$ symmetry
and not enhanced,
 can also easily be found.
These would be the analogues of the squashed $S^3$ example for the Type IX case above.

\section{An Extremal Brane Interpolating From Type VII to  $AdS_5$ }

So far, our emphasis has been on finding solutions which have a scaling symmetry along with the generalised translational 
symmetries discussed above. 
One expects  such scaling solutions to describe a dual field theory in the far infra-red and 
to  arise as the near-horizon limit of non-scale invariant solutions in general. 
An interesting feature of the field theory,  
 suggested by the geometric description, is that
in the  IR  it  effectively lives in a curved spacetime. 
For example, consider the scaling solution of Type VII type which was described in section \S\ref{simpletype7}.
 One expects, roughly, that the spacetime seen by the dual field theory is
given by a hypersurface at constant $r =r_0$ with $r_0\gg1$. 
From eq.(\ref{metricd}) we see that at $r=r_0$ this hypersurface is described by the metric, 
\be
\label{metricfield}
ds^2=R^2\left[-e^{2\beta_t r_0}dt^2+(dx^1)^2 + e^{2\beta r_0}[(\omega^2)^2+ \lambda^2 (\omega^3)^2]\right].
\ee
From eq. (\ref{defineomega}) we see that this is a non-trivial curved spacetime.

This observation raises a question: can the scaling solutions of non-trivial Bianchi type, 
 arise in situations 
where the field theory in the ultraviolet is in flat spacetime? Or do they require the field theory in the UV to 
itself be in a curved background? The latter case would be much less interesting from the
 point of view of possible connections with condensed matter systems. In the gravity description 
when the scaling solution arises from the near-horizon limit of a geometry 
which is asymptotically $AdS$ or Lifshitz, this  question takes the following form: 
is a  normalizable  deformation of the metric
(or its analogue in the Lifshitz case) enough? Or does one necessarily have to turn on a non-normalizable metric deformation
to obtain the scaling solution of non-trivial Bianchi type in the IR?

In this section we will examine this question for the Type VII solution discussed in section \S\ref{simpletype7}. 
We will find, after enlarging the   matter content to include two gauge fields, that  
normalizable deformations for the metric  suffice. 
More specifically, in this example we will find solutions  where the metric starts in the UV as being  $AdS$ space
with a normalizable metric perturbation being turned on and ends in the IR as a scaling Type VII solution. 
However, the gauge fields do have a non-normalizable mode turned on in the UV\footnote{One of the gauge
 fields corresponds to a current which is a relevant operator in the IR and the other to an irrelevant operator.
The current which breaks the symmetries of $AdS$ to those of Type VII is relevant in the IR.}. 
From the dual  field theory point of view  this corresponds to  being 
in flat space  but subjecting the system to 
a helically varying external current. In response,   the gravity solution teaches us that the  field theory
  flows to a  fixed point in the infra-red
with an ``emergent''  non-trivial metric of the Type VII  kind. 

The Type VII geometry breaks parity symmetry, as was discussed in section \S\ref{fivedim}. 
The Lagrangian for the example we consider preserves parity. In the solutions we find the parity violation 
is introduced by the non-normalizable deformations for the gauge fields, or in dual language,
by the external current source. 

In condensed matter physics  there are often situations where the Hamiltonian itself breaks parity
and the ground state of the system, in the absence of any external sources, 
 has a  helically varying  order parameter. This happens for example in the 
cases discussed  in \cite{condmatt1}, \cite{condmatt2}. It will be interesting to examine how such 
situations might arise in 
 gravitational systems as well. 
In $5$ dimensions  a Chern-Simons term can be written for a  gauge field and we expect that 
with parity violation  incorporated in  this way  gravity solutions showing similar behaviour can also be 
easily found.  For a  discussion of helical type instabilities along these lines see \cite{DH}, \cite{ooguri}, 
\cite{gauntlettspace}.

A similar analysis, asking whether a non-trivial asymptotic metric along the spatial directions
 is needed in the first place,  
 should also be carried out for  the other Bianchi classes. 
As a first pass one  can work in the thin wall approximation  and ask whether the  solution in the IR
with the required homogeneity symmetry can be matched to $AdS$ or a Lifshitz spacetime  using a domain wall
with  a tension that meets reasonable energy conditions. We leave this for future work.

\subsection{The System And The Essential Idea}
We will consider a system consisting of gravity with a cosmological constant 
and two gauge fields $A_1,A_2$, with masses, $m_1,m_2$ described by the Lagrangian
\be
\label{lagtwo}
S=\int d^5x \sqrt{-g}\{R + \Lambda -{1\over 4} F_1^2 -{1\over 4} F_2^2 -{1\over 4 } m_1^2 A_1^2 -{1\over 4} m_2^2
A_2^2\}.
\ee

We  show that this system has a  scaling solution of Type VII with parameters determined by 
$\Lambda,m_1^2,m_2^2$.  This scaling solution will be the IR end of the full solution we will construct which interpolates between the Type VII solution and asymptotic $AdS_5$ space in the UV. 

The essential idea in constructing this full solution is  as follows.
It will turn out that the scaling solution, as one parameter is made small, becomes \footnote{In the Type VII
 case  the symmetries are   generalised translations involving
 a combination of the usual translations and rotations along the spatial directions. In the  $AdS_2\times R^3$
case these symmetries 
 are enhanced to include both the usual translations and rotations along the  spatial directions. } 
$AdS_2\times R^3$.  We will call this parameter $\lambda$ below. 
Now it is well known that $AdS_2\times R^3$ arises as the near-horizon geometry of an extremal RN brane which asymptotes to $AdS_5$ space. 
This means the $AdS_2\times R^3$ solution  has  a perturbation\footnote{These perturbations are given in eq.(\ref{subleading1}) and
eq.(\ref{subleading2}) in \S\ref{Extremalrnbrane}.}
 which grows in the UV and takes it to $AdS_5$ space. 
Since the Type VII solution we find is close to $AdS_2\times R^3$ when $\lambda$ is small,
 it is easy to identify the corresponding perturbation in this solution as well. Starting with the Type VII
solution  in the IR  in the presence of this perturbation, 
we find, by numerically integrating the solution, that it  asymptotes to $AdS_5$ in the UV as well with only a 
normalizable perturbation for the metric being turned on. 

The Type VII solution, for small $\lambda$, is close to $AdS_2\times R^3$ in the following sense. 
At  values of the radial coordinate which are not too big in magnitude the Type VII solution becomes approximately 
$AdS_2\times R^3$ with a small perturbation being turned on. This means that when one starts with the Type VII solution in
 the IR, with the perturbation that takes it to $AdS_5$ being now turned on, the solution first evolves to being close to 
$AdS_2\times R^3$ at moderate values of $r$ and then further evolves to $AdS_5$ in the UV.

Below we  first describe the Type VII solution and the limit  where it becomes $AdS_2\times R^3$.
Then we  describe the extreme RN geometry and its near horizon $AdS_2\times R^3$ limit and  
finally we  identify the perturbation in the Type VII case and numerically integrate from the IR to UV showing that the
geometry asymptotes to $AdS_5$.

\subsection{Type VII Solutions}\label{Type VII geometry}
Let us  take the metric to be of the form
\be
\label{metricd2}
ds^2=R^2 \left\lbrace dr^2 -e^{2\beta_t r} dt^2+ (dx^1)^2 + e^{2\beta r} [(\omega^2)^2 + (\lambda+1) (\omega^3)^2] \right\rbrace.
\ee
This metric is the same as in eq.(\ref{metricd}) and has the same symmetries.
The only difference is that the parameter $\lambda$ has been redefined.
With our new definition  when $\lambda \rightarrow 0$ the metric becomes of Lifshitz type.
This follows from noting that $(\omega^2)^2+(\omega^3)^2 = (dx^2)^2+(dx^3)^2$, eq.(\ref{defineomega}).
In fact, once the equations of motion are imposed in this system, we will see that the $\lambda\rightarrow 0$ limit  gives 
$AdS_2\times R^3$. 
 
We take the gauge fields to have the form
\begin{eqnarray}
\label{tfgp2}
A_2 &=& e^{\beta r} \sqrt{\tilde{A_c}}\ \omega^2 \\
\label{tfgp3}
A_1 &=& e^{\beta_t r} \sqrt{\tilde{A_t}}\  dt.
\end{eqnarray}

Note that the $6$ variables, $R, \beta_t,\beta, \lambda, \tilde A_c,\tilde A_t,$ are all constants and  fully determine the
solution. We will now solve the equations of motion to determine these constants  in terms of the parameters
$\Lambda, m_1^2,m_2^2,$ which enter in the action eq.(\ref{lagtwo}).
As in the discussion of section \S\ref{fivedim} it will  be convenient to work in units where $R=1$. 

The gauge field equations of motion then  give
\begin{eqnarray}
\label{gfeom12}
 \sqrt{\ti A_t} \left[4 \beta \beta_t - m_1^2  \right] &=& 0\\
 \sqrt{\ti A_c} \left[2 \beta  (\beta_t + \beta)-\frac{2}{1+\lambda} -m_2^2   \right] &=& 0.
\end{eqnarray}

The trace reversed Einstein equations with components along $tt$, $rr$, $x^1x^1$
directions give:
\begin{eqnarray}
{2 \ti A_c \over 1 + \lambda} + 2 \ti A_c \beta^2 + \ti A_t (3 m_1^2 + 4 \beta_t^2 ) + 4 \left[ \Lambda - 3 \beta_t^2 - 6 \beta_t \beta \right] &=& 0\\
{\ti A_c \over 2(1 + \lambda)} - \ti A_c \beta^2 +\beta_t^2 ( \ti A_t-3) - 6 \beta^2 + \Lambda   &=& 0\\
{3 \lambda^2 + 2 \ti A_c \over 1 + \lambda } + \ti A_t \beta_t^2 - \ti A_c \beta^2 - 2 \Lambda &=& 0.
\end{eqnarray}
While the components with indices along the $x^2,x^3$ directions lead to
\begin{eqnarray}
{6 \lambda(2+\lambda)-2 \ti A_c  \over 1 + \lambda} + 3 m_2^2 \ti A_c + 4 \beta( \ti A_c \beta + 3 \beta_t + 6 \beta) + 2\ti A_t \beta_t^2 - 4 \Lambda &=&0\\
{3 \lambda (2 + \lambda ) - 2 \ti A_c  \over 1 + \lambda} + \ti A_c \beta^2 - \ti A_t \beta_t^2 - 6 \beta_t \beta - 12 \beta^2 + 2 \Lambda   &=&0.
\end{eqnarray}

These are $7$ equations in all, but one can  check that  only $6$ of them are independent and 
these $6$ independent equations determine the $6$ parameters mentioned above.
It turns out to be convenient to express  the resulting  solution as follows. 
Working in  $R=1$ units,  we 
 express the parameters in the Lagrangian, $\Lambda$ and $m_2^2$, 
 along with the constants in the solution $\beta_t,\beta,\ti A_t,\ti A_c$,  in terms of
$m_1^2$ and $\lambda$. This gives: 
\begin{eqnarray}
 m_2^2&=&-{2 \over 1+\lambda}+\lambda(2- \epsilon \lambda) \label{valm2}\\
\Lambda&=& {2 \over \epsilon} + 3 \epsilon \lambda^2+\lambda\left(-3+{\lambda \over 1-\lambda^2}\right) \label{vallambda}\\
\ti A_c&=&{2\lambda\over 1-\lambda}\\
\ti A_t&=&{4+\lambda\left(-4+\epsilon(5\lambda - 6)\right) \over 2(\lambda-1)(\epsilon \lambda-1)}\\
\beta_t&=&\sqrt{{2\over \epsilon}}(1-\epsilon \lambda)\\
\beta&=&\sqrt{{\epsilon \over 2}}\lambda,
\end{eqnarray}
where we have introduced a new parameter $\epsilon$ (not necessarily small) which is related to $m_1^2$ by 
\be
\label{defeps}
m_1^2= 4(1-\epsilon \lambda)\lambda.
\ee

Requiring $\ \ti A_t\ge0, \ \ti A_c \ge0,$ and also requiring that $\beta_t, \beta$ have the same sign,  leads to the 
conditions \footnote{The equations 
allow for $\beta_t,\beta$ to have opposite sign, this would correspond to 
a strange dispersion relation for fluctuations though and probably would be unstable. }
\be
\label{condlast}
0<\lambda <1, \ \ 0<\epsilon\le {4\over \lambda}{(1-\lambda)\over (6-5\lambda)}.
\ee
We have chosen both $\beta_t>0,\beta\ge0$, so that the horizon lies at $r\rightarrow -\infty$.

As in section \S\ref{fivedim}, we leave the important question about  the stability of these solutions for future work. 

\subsubsection{The Limit}
Now consider the limit where $\lambda=0$. In this limit, we get 
\begin{eqnarray}
\label{paralim}
m_2^2 = -2 \hspace{10mm} m_1^2=0 \hspace{10mm}  \ti A_c=0 \hspace{10mm} \beta =0 \\
\Lambda = {2 \over \epsilon} \hspace{10mm} \ti A_t = 2 \hspace{10mm} \beta_t =  \sqrt{{2\over \epsilon}}.
\end{eqnarray}
It is easy to see that the resulting geometry is $AdS_2\times R^3$. It is 
 sourced entirely by the gauge field $A_1$ which becomes massless. 
The second gauge field vanishes, since $\ti A_c=0$. 

Next consider a Type VII solution close to the limit $\lambda \rightarrow 0$, with $\epsilon$ being 
held fixed and of order unity.
In this limit,
\begin{eqnarray}
 \beta_t&=&\beta_t^{(0)}-\sqrt{2 \epsilon}\lambda \hspace{5mm} \text{where}\hspace{5mm}
\beta_t^{(0)}=\sqrt{{2\over \epsilon}} \label{tryone}\\
\beta &=& \sqrt{\epsilon \over 2} \lambda \label{trytwo} \\
\ti A_t&=&2-\epsilon\lambda+\mathcal{O}(\lambda^2) \label{trythree} \\
\ti A_c&=&2\lambda+\mathcal{O}(\lambda^2). \label{tryfour} 
\end{eqnarray}

Also we note that in this limit it follows from eq.(\ref{valm2}), eq.(\ref{vallambda}),  eq.(\ref{defeps})
 that 
\begin{eqnarray}
m_2^2 & = & -2+4 \lambda \label{smm2} \\
\Lambda & = & {2\over \epsilon}-3\lambda \label{smlambda} \\
m_1^2 & = & 4 \lambda \label{smm1}
\end{eqnarray}
up to $O(\lambda^2)$ corrections.

Now consider an intermediate range for the radial variable where
\be
\label{lima}
|r|\ll {1\over \lambda}.
\ee
In this intermediate region we can expand various components of the metric. E.g.,
\be
\label{exam}
e^{2\beta r} \simeq 1+ 2\beta r
\ee
etc.
This gives for the metric
\begin{eqnarray}
ds^2 & = & \left[dr^2-e^{2 \sqrt{\Lambda} r} dt^2 + (dx^1)^2+(dx^2)^2+(dx^3)^2\right]+ \delta ds^2 \label{appmet}
\\
A_1 & = & \sqrt{2} e^{\sqrt{\Lambda} r} dt + \delta A_1 \label{appa1} \\
A_2 & = & \delta A_2 \label{appa2}
\end{eqnarray}
where 
\begin{eqnarray}
\delta ds^2 & = & (\lambda r) \left[\sqrt{\epsilon\over 2}e^{2\sqrt{\Lambda}r} dt^2 + \sqrt{2\epsilon} ((dx^2)^2+(dx^3)^2)\right] + \lambda (\omega^3)^2 \label{appmet2} \\
\delta A_1 & = & -\left[\lambda r {\sqrt{\epsilon}\over 2} + {\lambda \epsilon \over \sqrt{8}} \right]
e^{\sqrt{\Lambda} r} dt \label{appa12} \\
\delta A_2 & = & \sqrt{2\lambda} \omega^2\label{appa22}.
\end{eqnarray}

In obtaining these expressions we have used  eq.(\ref{tryone}) and eq.(\ref{smlambda})
to express $\beta_t^{(0)}$ in terms of $\Lambda$.

We see that  when $\lambda \ll 1$ and eq.(\ref{lima}) is  met the solution does reduce to  $AdS_2\times R^3$ with only
$A_1$ being turned on. There are additional perturbations in the metric and gauge fields, given by 
eq.(\ref{appmet2}), eq.(\ref{appa12}), eq.(\ref{appa22}). These are  small in this region.

\subsection{Extremal RN Brane}\label{Extremalrnbrane}
In this subsection we turn off the gauge field $A_2$ and consider a system  consisting of just gravity, 
with $\Lambda$, and  $A_1$, with no mass. 
It is well known that this system has an extremal Reissner Nordstrom black brane solution
with metric
\begin{equation}
 ds^2=-a(\tir)dt^2 + {1\over a(\tir)} d\tir^2 +b(\tir)[dx^2+dy^2+dz^2]
\end{equation}
where 
\be
\label{metern}
a(\tir)={(\tir - \tir_h)^2 (\tir + \tir_h)^2 (\tir^2 + 2 \tir_h^2) \Lambda\over 12 \tir^4} ;
\hspace{10mm}  b(\tir)=\tir^2
\ee
and gauge field
\be
\label{gfern}
A=A_t dt
\ee
where 
\be
\label{gfern2}
A_t(\tir)= -\tir_h \sqrt{\frac{\Lambda}{2}} \left(\frac{\tir_h^2}{\tir^2}-1\right).
\ee
The parameter $\tir_h$ determines the location of the horizon and is determined by the charge density carried by the extremal brane. 

Near the horizon, as $\tir \rightarrow \tir_h$, the geometry becomes of $AdS_2\times R^3$ type. 
As discussed in Appendix \ref{AppD} by changing variables
\begin{equation}
\label{defrtilde}
 r=\int \frac{d\tir}{\sqrt{a(\tir)}}
\end{equation}
one can express the metric and gauge fields for $\tir \rightarrow \tir_h$ as 
\begin{eqnarray}
 ds^2 &=&  dr^2-  e^{2\sqrt{\Lambda} r} \left(1- \frac{14}{3}e^{\sqrt{\Lambda} r}\right)dt^2+ \left(1+2 e^{\sqrt{\Lambda} r}\right) (dx^2+dy^2+dz^2) \label{metint} \\
A_t &=&  \sqrt{2}e^{ \sqrt{\Lambda} r}\left(1-\frac{8}{3} e^{\sqrt{\Lambda} r} +\cdots\right) dt \label{gint}.
\end{eqnarray}

It is useful to write these expressions as the   leading terms
\begin{eqnarray}
\label{leading}
ds^2 & = & dr^2-e^{2\sqrt{\Lambda} r} dt^2 +(dx^1)^2+(dx^2)^2+(dx^3)^2\\
A_{1t} & = & \sqrt{2} e^{\sqrt{\Lambda}r},
\end{eqnarray}
which correspond to $AdS_2\times R^3$ sourced by a gauge field in the time-like direction,
and corrections, 
\begin{eqnarray}
\delta ds^2 &=&  \frac{14}{3}e^{3\sqrt{\Lambda} r}dt^2
+2e^{\sqrt{\Lambda}r}\left((dx^1)^2 +(dx^2)^2+(dx^3)^2\right) 
\label{subleading1}\\
\delta A_{1t} & = & -{8\over 3}  \sqrt{2} e^{2 \sqrt{\Lambda} r} 
\label{subleading2}.
\end{eqnarray}
We see that the perturbations are small if 
\be
\label{condro}
e^{\sqrt{\Lambda} r} \ll 1,
\ee
which is the condition for being close enough to the horizon of the extremal brane. 
Note that this condition requires that $r$ be  negative.

% The perturbation in eq.(\ref{subleading1}) and eq.(\ref{subleading2}) promote 
%  $AdS_2 \times R^3$ at IR into $AdS_5$ at UV. Therefore we will now consider adding this perturbation to 
%  type VII solution, which is close to $AdS_2 \times R^3$ at small $\lambda$.

\subsection{The Perturbation in Type VII}
We can now  identify the perturbation in Type VII, with small $\lambda$, which becomes
eq.(\ref{subleading1}), eq.(\ref{subleading2}) in the $\lambda\rightarrow 0$ limit. 
We write
\begin{eqnarray}
ds^2 & = & dr^2-f(r) dt^2 + g(r) (dx^1)^2 + h(r) (\omega^2)^2 + j(r) (\omega^3)^2 \label{fmet} \\
A_1 & = & A_{1t}(r) dt \label{fa1}\\
A_2 & = & A_2(r) \omega^2, \label{fa2}
\end{eqnarray}
with
\begin{eqnarray}
f(r) & = & e^{2\beta_t r}(1 + \tilde{\epsilon}  f_c e^{\nu r} ) \\
g(r) & = & 1+ \tilde{\epsilon} g_c e^{\nu r} \\
h(r) & = & e^{2\beta r}(1+ \tilde{\epsilon} h_c e^{\nu r}) \\
j(r) & = & e^{2\beta r} \left( \lambda + 1 \right)  \left( 1+ \tilde{\epsilon} j_c e^{\nu r} \right) \\
A_{1t}(r) & = & \sqrt{\tilde{A}_t} e^{\beta_t r} ( 1+ \tilde{\epsilon} A_{1c} e^{\nu r}) \\
A_{2}(r) & = & \sqrt{\tilde{A}_c}e^{\beta r}(1+ \tilde{\epsilon} A_{2c} e^{\nu r}).
\end{eqnarray}
The parameters $\beta_t,\beta, A_{1c}, A_{2c}$ take values given in eq.(\ref{tryone}), eq.(\ref{trytwo}), eq.(\ref{trythree}), eq.(\ref{tryfour}), and  
$\tilde{\epsilon}$ is the small parameter which keeps the perturbation small. 

It is easy to see that there is a  solution to the resulting coupled linearized equations in which 
\be
\label{valpnu}
\nu=\sqrt{2\over \epsilon}\left(1 - \lambda {20\over 9} \epsilon \right).
\ee
By shifting $r$ we can always rescale  $f_c$.  
For comparing with eq.(\ref{subleading1}), eq.(\ref{subleading2}) we take 
\be
\label{valepma}
 \tilde{\epsilon} f_c= -{14\over 3}.
\ee 
 The other parameters then take the values   
\begin{eqnarray}
 \ti \epsilon g_c&=&2\left(1+\lambda \ {\epsilon\over378}(167+54\epsilon)\right)\\
\ti \epsilon h_c&=&2\left(1+\lambda \ {\epsilon(548-\epsilon(89+81\epsilon))\over 756(\epsilon-1)}\right)\\
\ti \epsilon j_c&=&2\left(1+\lambda \ {\epsilon(170-\epsilon(467+81\epsilon))\over 756(\epsilon-1)}\right)\\
\ti \epsilon A_{1c}&=& -{8\over3}\left(1+\lambda \ {(32-27\epsilon)\epsilon \over2016}\right) \\
\ti \epsilon A_{2c} & = & -{1\over 2} \epsilon  + O(\lambda).
\end{eqnarray}
Note the $O(\lambda)$ correction to $A_{2c}$ will contribute at higher order since $\tilde{A}_c$ is or order 
$\lambda$. 

It is easy to see that as $\lambda \rightarrow 0$ this perturbation becomes exactly the one given in 
eq.(\ref{subleading1}), eq.(\ref{subleading2}).

\subsection{The Interpolation}

We are finally ready to consider what happens if we start with the Type VII solution we have found 
in section \S\ref{Type VII geometry} but now with the perturbation identified in the previous section being turned on. 
We work with the cases where $\lambda \ll 1$. 

Since $\nu>0$ we see that as $r\rightarrow -\infty$ the effects of the perturbation becomes very small and the solutions becomes the Type VII solution  discussed in section \S\ref{Type VII geometry}. 

In the region where $r$ is negative and both conditions eq.(\ref{lima}) and eq.(\ref{condro})
 are met we see that the resulting 
solution can be thought of as 
 being approximately $AdS_2\times R^3$ with only $A_1$ being turned on along the time direction. There are corrections to this approximate solution given by the sum of the perturbations in
 eq.(\ref{appmet2}), eq.(\ref{appa12}),eq.(\ref{appa22}) and eq.(\ref{subleading1}), eq.(\ref{subleading2}).   

To numerically integrate we will start in this region and then go further out to larger values of $r$. 
We take the metric and gauge fields to be  given by eq.(\ref{fmet}), eq.(\ref{fa1}), eq.(\ref{fa2}). 

We are interested in a solution which asymptotes to $AdS_5$. It is best to work with values of 
$\epsilon$ for which $m_2^2$  lies above the $AdS_5$ BF bound. 
This condition gives
$m_2^2/\Lambda>-{1\over 6}$. Working with the leading order terms in eq.(\ref{smm2}), eq.(\ref{smlambda}) this condition is met if
\be
\label{condepsa}
\epsilon<{1\over 6}.
\ee

In the figures which are included below we have taken $\epsilon={1\over 7}$.
In addition, we take $\lambda = 10^{-2}$.
And  we start the numerical integration at $r=-3$, which meets the conditions
eq.(\ref{lima}), eq.(\ref{condro}). Varying these parameters,  within a range of values, does not change 
our results in any essential way.
\begin{figure}
 \begin{center}
 \includegraphics[scale=0.9]{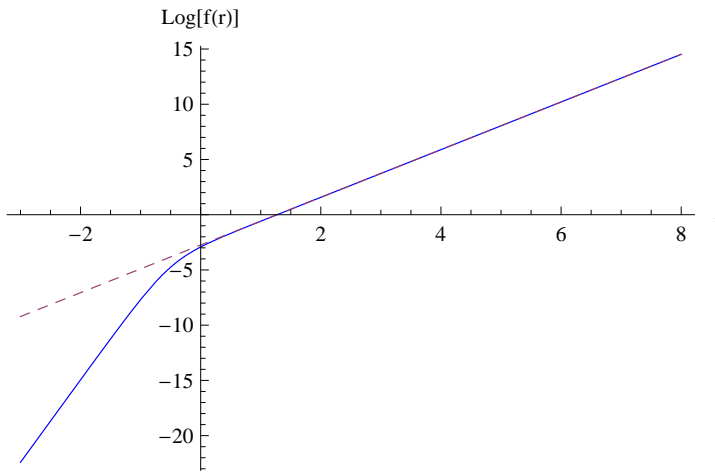}
 \caption{Numerical solution for metric component $\log f(r)$.}
 \end{center}
 \end{figure}
  
 \begin{figure}
 \begin{center}
 \includegraphics[scale=0.9]{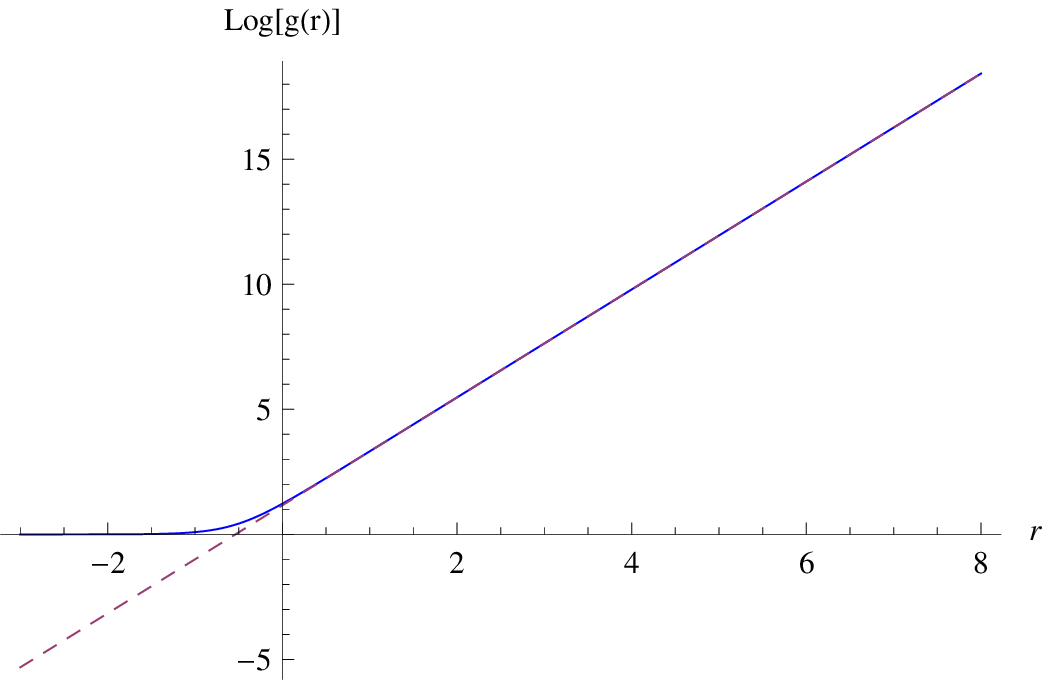}
 \caption{Numerical solution for metric component $\log g(r)$.}
 \end{center}
 \end{figure}
  
 \begin{figure}
 \begin{center}
 \includegraphics[scale=0.9]{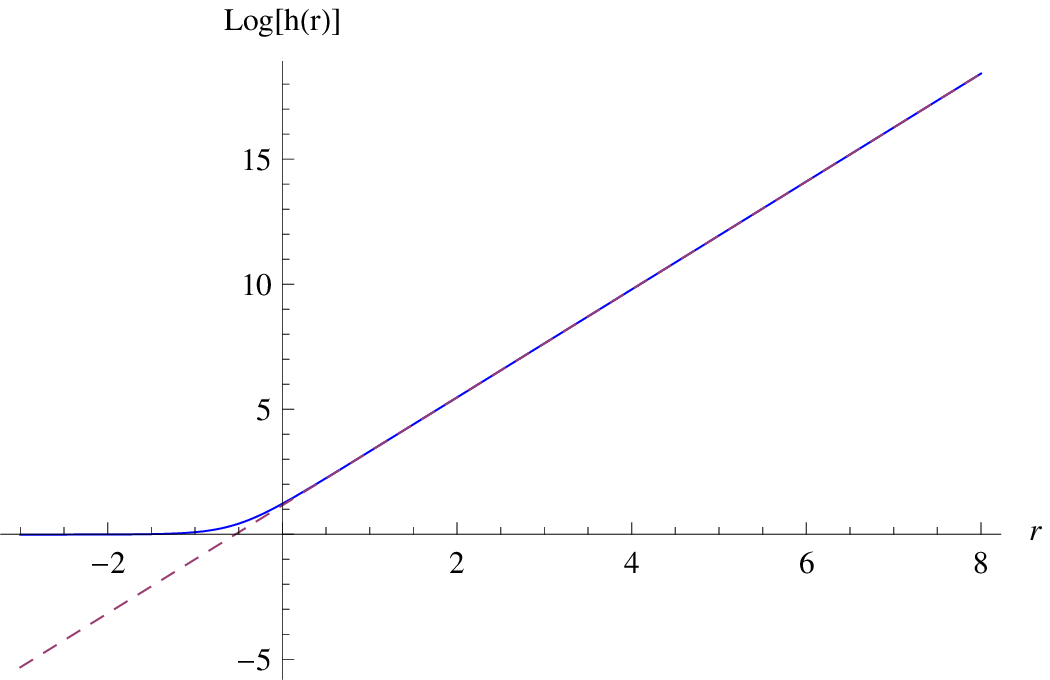}
 \caption{Numerical solution for metric component $\log h(r)$.}
 \end{center}
 \end{figure}
  
 \begin{figure}
 \begin{center}
 \includegraphics[scale=0.9]{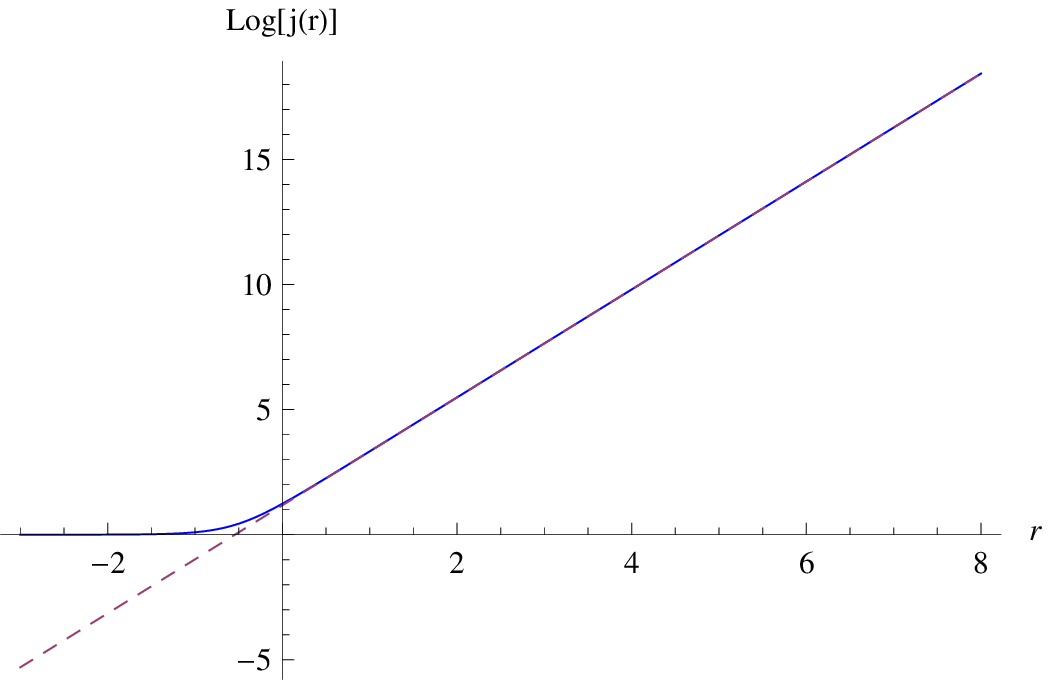}
 \caption{Numerical solution for metric component $\log j(r)$.}
 \end{center}
 \end{figure}

 \begin{figure}
 \begin{center}
 \includegraphics[scale=0.9]{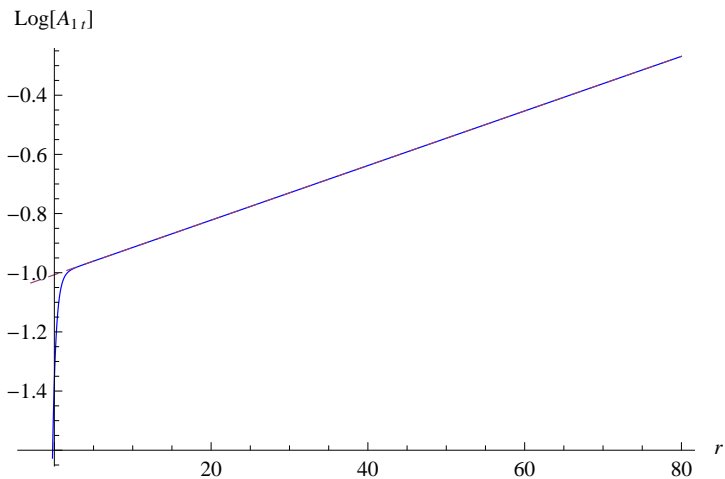}
 \caption{Numerical solution for gauge field  $\log A_{1t}(r)$.}
 \end{center}
 \end{figure}
 
 \begin{figure}
 \begin{center}
 \includegraphics[scale=0.9]{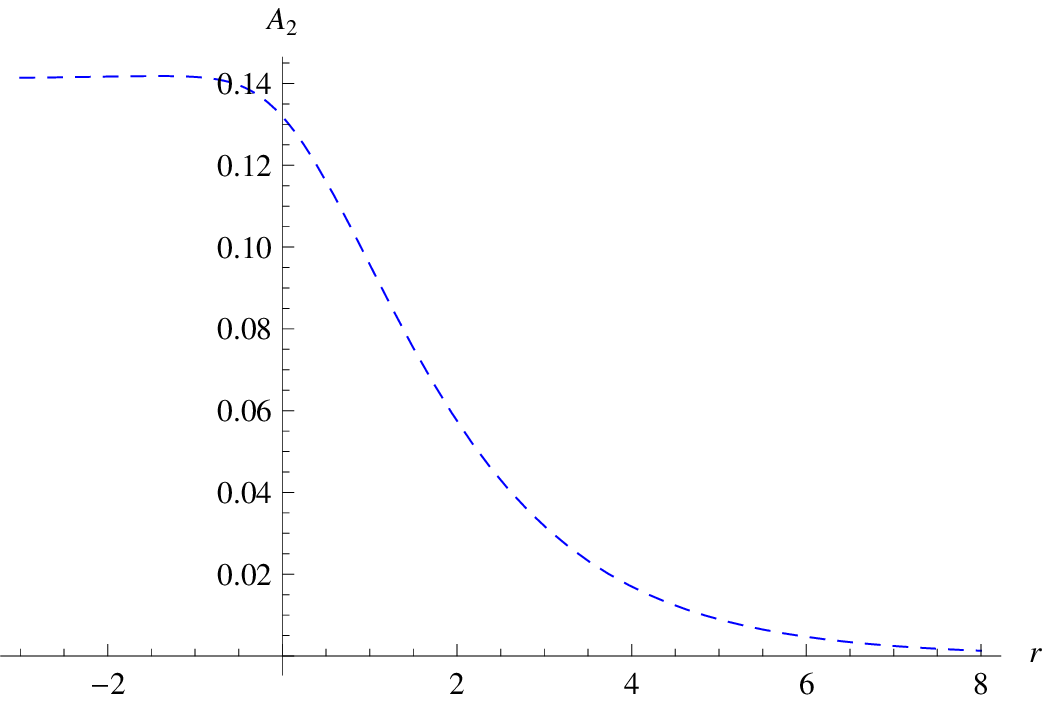}
 \caption{Numerical solution for gauge field  $ A_{2}(r)$.}
 \end{center}
 \end{figure}
The reader will see from Fig.1, 2, 3, 4,  that the metric coefficients $f(r), g(r), h(r), j(r),$ 
become exponential in $r$ as $r\rightarrow \infty$. We have fitted the slope and find agreement with the behavior in $AdS$ space with cosmological constant given in eq.(\ref{smlambda}). 

Notice that  when $\lambda$ is small and positive the gauge field $A_1$ has a small and positive $m^2$, 
eq.(\ref{smm1}). 
In asymptotic $AdS_5$ space such a gauge field has two solutions behaving like 
\be
\label{twosola1}
A_{1} \sim e^{ \left(-\sqrt{\Lambda \over 12} \pm \sqrt{{\Lambda \over 12}+{m_1^2 \over 2}} \right) r}.
\ee
In particular one  branch grows exponentially  as $r\rightarrow \infty$. It is easy to see that this behavior agrees with the Fig. 5 at large $r$. The small value of $m_1^2$ means one has to go out to relatively large values of $r$ to see this behavior. 
In contrast, since $A_2$ has  negative  $m^2$ both the solutions asymptotically die out, which agrees
with Fig. 6.
 
One comment is worth making here. 
The reader might wonder why the asymptotically growing $A_1$ gauge field does not back react and lead to 
a departure from $AdS_5$ space. The reason is that  $m_1^2$ is very slightly positive, {\it i.e.,} $m_1^2/\Lambda \ll 1$, and as a result the 
stress energy due to this field is still subdominant compared to the cosmological term at large $r$. 

In summary,
 the numerical integration shows that the resulting geometry is asymptotically $AdS_5$ with small 
corrections which correspond to normalizable perturbations being turned on. 
The gauge fields in contrast have both their normalizable and non-normalizable modes turned on. 
This is clear in the $A_1$ case which grows exponentially, but it is also true for $A_2$ which has 
negative $m^2$ so that both modes for it die away asymptotically. 

We have thus established that the Type VII geometry discussed in section \S\ref{Type VII geometry} can arise starting from a flat boundary metric with non-normalizable gauge field deformations turned on.

\section{Generalised Translations Involving Time and  Closed Time-Like Curves}
So far, we have been discussing examples where the usual time translation symmetry is preserved and the generalised translations 
only involve the spatial directions. However, one can  
also have situations   where the generalised translations  involve  the time direction in a non-trivial manner. 
For example,  a case  of Type $VII_0$ type  can 
  arise when a translation in the  time direction is accompanied by a rotation 
in a spatial two-plane. This could happen 
  if there is a vector order parameter which is  time dependent and 
rotates in the spatial two-plane as time advances.  The time dependent order parameter might arise as a response 
to a suitable periodic time dependent external source. 

We should note that a situation where time gets non-trivially involved in the 
 generalised translations   does not 
necessarily require the spacetime to be time dependent.
For example it could have been that one has the standard translational symmetry along time and all the 
 spatial directions except for one, $x^1$, and in the $x^1$  direction the  translation is accompanied by a boost in the 
$t,x^2$ plane. 
In this case the metric would depend on $x^1$ but not $t$, just as in the case of Type VII
discussed in section \S\ref{simpletype7} the metric depended on $x^1$ but not $x^2,x^3$. 
   Another  example will be presented below where this point will also become clear.

Here, as in the discussion earlier, we will be interested in geometries which have scale invariance, along with the generalised translational symmetries mentioned above. 
Such gravitational solutions, from the point of view of possible connections with condensed matter physics, could describe 
dynamical critical phenomenon which arise  when a   system is subjected to  time varying external forces. 

In general, in the $9$ Bianchi classes, 
the three generators of the symmetry algebra are inequivalent. Once we allow for the generalised translations
  to also involve time one  typically finds that
each Bianchi class gives rise to more than one physically distinct possibility
depending on the role time translations play in the symmetry algebra.

We have not systematically examined which of the many resulting cases   arising in this way 
can be realized by coupling reasonable kinds of 
matter fields to gravity, or  even examined systematically which of these cases are   allowed by the various energy conditions.
These conditions   could well  impose fairly  stringent restrictions in these cases where time is more non-trivially involved.

A few simple examples which we have constructed already  illustrate some  novel aspects which arise in 
such situations  and we will limit ourselves to commenting on them here.  
In particular  we will discuss below  one example in which the resulting space-time may be physically 
unacceptable since it involves closed time-like curves. 
This example illustrates the  need to proceed with  extra caution in dealing with situations where the generalised
translations also involve time.

The case we have in mind is most simply discussed in $4$ total spacetime dimensions and 
is based on the Bianchi Type II algebra, which involves translations of a generalised sort along  $t,x^1,x^2$, generated by 
Killing vectors
\be
\label{killvec}
\xi_1=\partial_t, \ \  \xi_2=\partial_1, \ \ \xi_3=\partial_2+x^1 \partial_t. 
\ee
The invariant one-forms are given by (see Appendix A)
\begin{equation}\label{tdefomega}
 \omega^1= dt-x^2 dx^1 \ ; \ \omega^2=dx^1  \ ; \ \omega^3=dx^2.
\end{equation}
We assume the full metric to be also invariant under 
a translation in $r$ accompanied by a rescaling of $t,x^1,x^2$. 
Also we will take the metric to be  diagonal in the basis of the invariant one-forms given above which gives 
\footnote{Note that this metric is not time dependent, 
also note that for simplicity we are assuming that both spatial coordinates $x^1$ and $x^2$ scale similarly, but in general that is not required.}
\be
\label{metricd3}
ds^2=R^2\left[dr^2 -e^{2 \beta_1 r} (dt-x^2 dx^1)^2+ e^{\beta_1 r }((dx^1)^2 + (dx^2)^2) \right].
\ee

The gauge field is taken to be along the invariant one-form $\omega^1$ given in eq.(\ref{tdefomega}) and of the form
\be
\label{gft}
A= \sqrt{A_1} e^{\beta_1 r} \omega^1.
\ee

The gauge field equation of motion for this system gives
\be
m^2 - 2 (1 + \beta_1^2)=0.
\ee
The  trace reversed Einstein equations along $r,t,x^1$ respectively are
\begin{eqnarray}
  A_1 - 6 \beta_1^2 + A_1 \beta_1^2 + 2 \Lambda&=&0\\
-2 + A_1 + A_1 m^2 - 8 \beta_1^2 + A_1 \beta_1^2 +  2 \Lambda&=&0\\
2 - A_1 - 4 \beta_1^2 - A_1 \beta_1^2 + 2 \Lambda&=&0.
\end{eqnarray}
The Einstein equation along $x^2$ gives the same equation as along $x^1$.

It can be easily verified that the solution to these equations is
\begin{equation}
 m^2 = 2 (1 + \beta_1^2) \hspace{10mm} \Lambda = {1 \over 2 } (5 \beta_1^2 -1) \hspace{10mm} A_1 =1.
\end{equation}

So we see that reasonable matter in the form of a massive Abelian gauge field,
 in the presence of a cosmological constant, can 
give rise to a geometry of Type II where time enters in a non-trivial way in the generalised translations. 

However, as we will now see, this metric has closed time-like curves. This is a cause for physical concern, although
similar solutions were investigated in \cite{DeWolfe} and suggestions about how CTCs may sometimes be acceptable 
features in such solutions were described there. 

With a  redefinition of the form,
\begin{equation}
t-{x^1x^2\over 2}\rightarrow t\hspace{5mm} ; \hspace{5mm} x^1\rightarrow\rho
\sin{\theta}\hspace{5mm} ; \hspace{5mm}
x^2\rightarrow\rho \cos{\theta}
\end{equation}
the metric becomes
\begin{equation}
 ds^2=dr^2 - e^{2 \beta_1 r} dt^2 +  e^{\beta_1 r} d\rho^2 + e^{\beta_1 r} \rho^2\left(1 -{\rho^2\over 4} e^{ \beta_1 r} \right) d\theta^2 +\rho^2 e^{2 \beta_1 r} d\theta dt.
\end{equation}
Here the coordinate $\theta$ is periodic with period $2\pi$. 
Now notice that   for $\left(1 -{\rho^2\over 4} e^{ \beta_1 r}\right)<0$, $\theta$   becomes time-like and
 the closed curve, $\theta=[0, 2\pi]$,  becomes  a closed time-like curve \footnote{This observation was made
by R. Loganayagam. We thank him for some very helpful conversations.}.

\section{Discussion}
In this paper, we argued that black branes  need not be translationally invariant along 
the spatial directions in which they extend, and could instead have the less restrictive property of 
homogeneity. We showed that such  black brane solutions in 
$4+1$ dimensions are classified by the Bianchi classification and fall into $9$ classes. 
We mostly considered  extremal black branes and focused  on their near-horizon  
attractor region,
which we took to be of  scale invariant type. 
We  found that such scale invariant solutions, realizing many of  the non-trivial Bianchi classes,
can arise   in  relatively simple gravitational systems with a negative cosmological constant and  
 one or two massive Abelian gauge fields as matter. 
From the point of view of the holographically dual field theory, and of possible connections with condensed matter physics,
these  solutions correspond to new phases of matter which can arise when translational invariance is 
broken\footnote{Massive gauge fields in the bulk correspond to  currents in the boundary theory which are
 not conserved. One  expects that similar solutions should 
 also arises when the symmetry is spontaneously broken in the
bulk, corresponding to a superfluid in the boundary. It will be interesting to check if superfluids found in nature exhibit all of the phases we have found.}.

It is clear that this paper has merely scratched the surface in exploring 
 these    new  homogeneous brane solutions, and  much more needs to be done to understand 
them  better. Some directions for future work are as follows:

\medskip
\noindent
$\bullet$
One would like  to analyze the stability properties of the near-horizon solutions we have found 
in more detail.  In the  examples we constructed, sometimes solutions in two different classes are allowed 
for the same range of parameters. 
For example, with one gauge field,
both Lifshitz and Type VII can arise for the same mass range. In such cases especially one would like to 
know if both  solutions are stable or if one of them  has an instability. 

\medskip
\noindent
$\bullet$
It seems reasonable to suggest that the scaling solutions we have obtained arise as the near-horizon 
limits of extremal black brane solutions.  For the case of Type VII we showed that this was indeed true
by obtaining  a full extremal black brane solution which asymptotes to $AdS$ space  at large distance. 
 It would  be worthwhile to check if this is true
for the other classes as well \footnote{In particular it is worth checking whether the asymptotic geometry
along the spatial directions can be taken to be flat space for the other classes as well, 
as happens in the Type VII example we have constructed. We expect this to be true  in several cases, for suitable
matter content.}.

\medskip
\noindent
$\bullet$
One would like to find the small temperature deformation of these solutions. This can be done working
in the near-horizon limit itself. 
This is important in characterizing the thermodynamics of these systems better, and also in establishing that the solutions are the end points of non-extremal solutions. 
It is known that while curvature invariants are finite in the Lifshitz solutions, tidal forces do blow up (for a recent discussion of the nature of Lifshitz singularities, see e.g. \cite{horowitz2}).
We saw that this was also true in the Type VII case investigated in \S\ref{conrem}, and expect it to be true in many  solutions in the other Bianchi classes as well.
It would be  worth investigating this issue  further. 
If a finite temperature deformation exists, it would probably allow us to control 
 the effects due to these large tidal forces. 

\medskip
\noindent
$\bullet$
One should  understand these solutions more deeply from the point of view of
the attractor mechanism. The solutions are tightly constrained by  symmetries,
since they have the  generalised translational invariances and scaling invariance (this involves a
 translation along the radial direction). These symmetries are enough to determine the
solution completely up to a few constants, e.g. $R,\beta_1,\beta_i, A_t$ in eq.(\ref{metricc}), eq.(\ref{fgp}).
It should be straightforward to formulate an entropy function for any given Bianchi class
which can be extremised to obtain these constants,
thereby  determining the full solution from purely algebraic considerations \footnote{We are thankful to Ashoke Sen for very helpful discussions
in this regard.}. 
It would also be interesting to ask when a given scaling solution is an attractor in terms of varying the
asymptotic data and studying how the solution changes, and to understand how to encode the attractor behavior
(or lack thereof) in terms of  suitable near-horizon data. We hope to report on this development soon. 

\medskip
\noindent
$\bullet$
Going beyond solutions with scaling symmetry, one would  like to incorporate the   possibility that 
the near-horizon geometry does not have a scaling isometry, but instead has a conformal Killing vector
corresponding to a shift of the radial direction and a rescaling of the other directions.  
Such metrics have been analyzed in \cite{kiritsis}, \cite{perlmutter}, \cite{iknt}, \cite{kiritsismore}, 
\cite{takayanagi}, \cite{subir}, \cite{Dong}
and shown to lead to interesting behaviours. 

\medskip
\noindent
$\bullet$
It is quite amazing to us that even the relatively simple matter systems  we have explored here,
 consisting of a just a couple of massive  gauge fields, can yield such a large diversity of  solutions.
 It would be fascinating to explore  within  a simple system of this type,
which of the  solutions are related to each other 
by RG flow. 

\medskip
\noindent
$\bullet$
Naturally,  one would like to know  whether the kinds of simple gravitational systems we have analyzed here
(with their required values of parameters) can arise in string theory\footnote{Some brane solutions of supergravity theories with 
instabilities which break translational invariance  have been 
identified in the literature \cite{ooguri}, \cite{gauntlettspace}. 
It would also  be interesting to ask whether the attractor solutions we have found  describe the
 end points for these instabilities.}.
To be even more ambitious, one would ultimately like to ask about all of the phases 
that can be realized 
in gravitational systems which admit a consistent UV completion. 
The homogeneous phases we have investigated here are many more than were previously considered, 
but one suspects that even they  are only  a tiny subset of all possibilities! 

Clearly, in gravity, as in condensed matter physics,  more is different and  there is much left to be done.

\newpage

\bigskip
\centerline{\bf{Acknowledgements}}
\medskip

We thank  Gary Horowitz, Renata Kallosh, R.~Loganayagam, Kengo Maeda, Gautam ~Mandal, Shiraz ~Minwalla,
Mike Mulligan,  Eric Perlmutter, Shiroman ~Prakash, Ramadas Ramakrishnan 
  and  especially Kedar ~Damle and Ashoke ~Sen  for helpful discussions.
NI, SK and SPT thank the organizers of the meeting on ``Hot Nuclei, Cold Atoms and Black Holes'' 
held at the ASICTP, Trieste.  SPT also thanks the organizers of the  Higher Spin Field Theories meeting held in HRI, Allahabad, 
 and  NK, PN,  NS and SPT thank the organizers of the National Strings Meeting held  Delhi, for 
 providing a stimulating atmosphere where  some of this research was carried out. 
 NI would like to thank RIKEN for their kind hospitality where part of this work was done. 
 SK acknowledges the Aspen Center for Physics for hospitality while this work was in 
 progress, and
 thanks the organizers of the  ``Holographic Duality and
 Condensed Matter Physics" workshop at the KITP for providing such a stimulating atmosphere.
  NK, PN, NS  and SPT acknowledge funding from the Government of India,
and thank the people of India for  generously supporting research in string theory.
SK was supported by the US DOE under contract DE-AC02-76SF00515 and the
National Science Foundation under grant no. PHY-0756174.

\newpage

\appendix

\section{Three Dimensional Homogeneous spaces}\label{3dhomogspace}\label{AppA}
A three dimensional homogeneous space has three linearly independent Killing vector fields, $\xi_i$, $i=1,2,3$.
The infinitesimal transformations generated by these Killing vectors take any point in the space
 to any other point in 
its immediate vicinity. The  Killing vectors satisfy a three dimensional real  algebra with commutation relations
\begin{equation}
\left[\xi_i,\xi_j\right] = C^k_{ij} \xi_k.
\end{equation}
As discussed in \cite{LL}, \cite{SR} there are $9$ different such algebras, up to basis redefinitions, 
and thus $9$ different classes of such homogeneous spaces. 
This classification is called the Bianchi classification and the $9$ classes are the $9$ Bianchi classes. 

Also, as is discussed in \cite{LL},\cite{SR}, in each case there are three linearly independent 
invariant vector fields, $X_i$, which commute with the three Killing vectors
\be
\label{comm}
[\xi_i,X_j]=0.
\ee
The $X_i$'s in turn satisfy the algebra
\be
\label{alinv}
[X_i,X_j]=-C^k_{ij}X_k.
\ee
There are also three one-forms, $\omega^i$, which are dual to the invariant vectors $X_i$.
 The Lie derivatives of 
these one-forms along the $\xi_i$ directions also vanish making them invariant along the $\xi$ directions
as well. 
The $\omega^i$'s satisfy the relations
\be
\label{diffonef}
d\omega^i = \frac{1}{2} C^i_{jk} \omega^j \wedge \omega^k.
\ee

Below we give a list which contains  the structure constants for the $9$ Bianchi algebras,
 in a particular basis of generators.
For all the classes that arise  in this paper we also give the 
Killing vector fields, invariant one-forms and invariant vector fields,  in a particular coordinate 
basis. See \cite{SR} for more details.

\begin{itemize}
 \item {\bf Type I}: $C^i_{jk}=0$\\
\begin{equation}
 \xi_i=X_i=\partial_i,~\omega^i=dx^i,~d\omega^i=0
\end{equation}

\item {\bf Type II}: $C^1_{23}=-C^1_{32}=1$ and rest $C^{i}_{j,k}=0$\\
\begin{tabular}{cccc}
$\xi_1=\p_2$ & $X_1=\p_2$ & $\omega^1=dx^2-x^1 dx^3$ &
$d\omega^1=\omega^2\wedge\omega^3$ \\
$\xi_2=\p_3$ & $X_2=x^1\p_2+\p_3$ & $\omega^2=dx^3$ & $d\omega^2=0$ \\
$\xi_3=\p_1+x^3 \p_2$ & $X_3=\p_1$ & $\omega^3=dx^1$ & $d\omega^3=0$
\end{tabular}

\item {\bf Type III}: $C^1_{13}=-C^1_{31}=1$ and rest $C^{i}_{j,k}=0$\\
\begin{tabular}{cccc}
$\xi_1=\p_2$ & $X_1=e^{x^1}\p_2$ & $\omega^1=e^{-x^1}dx^2$ &
$d\omega^1=\omega^1\wedge\omega^3$ \\
$\xi_2=\p_3$ & $X_2=\p_3$ & $\omega^2=dx^3$ & $d\omega^2=0$ \\
$\xi_3=\p_1+x^2 \p_2$ & $X_3=\p_1$ & $\omega^3=dx^1$ & $d\omega^3=0$
\end{tabular}

\item {\bf Type V}: $C^1_{13}=-C^1_{31}=1$, $C^2_{23}=-C^2_{32}=1$  and rest $C^{i}_{j,k}=0$\\
\begin{tabular}{cccc}
$\xi_1=\p_2$ & $X_1=e^{x^1}\p_2$ & $\omega^1=e^{-x^1}dx^2$ &
$d\omega^1=\omega^1\wedge\omega^3$ \\
$\xi_2=\p_3$ & $X_2=e^{x^1}\p_3$ & $\omega^2=e^{-x^1}dx^3$ & $d\omega^2=\omega^2\wedge\omega^3$ \\
$\xi_3=\p_1+x^2 \p_2+ x^3 \p_3$ & $X_3=\p_1$ & $\omega^3=dx^1$ & $d\omega^3=0$
\end{tabular}

\item {\bf Type VI}: $C^1_{13}=-C^1_{31}=1$, $C^2_{23}=-C^2_{32}=h$ with ($h\ne0,1$) and rest $C^{i}_{j,k}=0$\\
\begin{tabular}{cccc}
$\xi_1=\p_2$ & $X_1=e^{x^1}\p_2$ & $\omega^1=e^{-x^1}dx^2$ &
$d\omega^1=\omega^1\wedge\omega^3$ \\
$\xi_2=\p_3$ & $X_2=e^{h x^1}\p_3$ & $\omega^2=e^{-h x^1}dx^3$ & $d\omega^2=h \omega^2\wedge\omega^3$ \\
$\xi_3=\p_1+x^2 \p_2+h x^3 \p_3$ & $X_3=\p_1$ & $\omega^3=dx^1$ & $d\omega^3=0$
\end{tabular}

\item {\bf Type $\bf{VII_0}$}: $C^1_{23}=-C^1_{32}=-1$, $C^2_{13}=-C^2_{31}=1$ and

rest $C^{i}_{j,k}=0$.\\
\begin{tabular}{cccc}
$\xi_1 = \partial_2$ & $X_1=\cos(x^1)\p_2+\sin(x^1)\p_3$\\
$\xi_2 = \p_3$ & $X_2=-\sin(x^1)\p_2+\cos(x^1)\p_3$\\
$\xi_3=\p_1-x^3 \p_2+x^2 \p_3$ &  $X_3=\p_1$
\end{tabular}

And also,

\begin{tabular}{cc}
$\omega^1=\cos(x^1)dx^2+\sin(x^1)dx^3$ & $d\omega^1=-\omega^2\wedge\omega^3$ \\
$\omega^2=-\sin(x^1)dx^2+\cos(x^1)dx^3$ & $d\omega^2=\omega^1\wedge\omega^3$ \\
$\omega^3=dx^1$ & $d\omega^3=0$
\end{tabular}

\item {\bf Type IX}: $C^1_{23}=-C^1_{32}=1$, $C^2_{31}=-C^2_{13}=1$, $C^3_{12}=-C^3_{21}=1$ and rest are zero.

\begin{tabular}{cccc}
$\xi_1 = \partial_2$\\
$\xi_2 = \cos(x^2)\p_1-\cot(x^1)\sin(x^2)\p_2+{\sin(x^2)\over \sin(x^1)}\p_3$\\
$\xi_3=-\sin(x^2)\p_1-\cot(x^1)\cos(x^2)\p_2+{\cos(x^2)\over \sin(x^1)}\p_3$
\end{tabular}

With

\begin{tabular}{cccc}
$X_1=-\sin(x^3)\p_1+{\cos(x^3)\over\sin(x^1)}\p_2-\cot(x^1)\cos(x^3)\p_3$\\
$X_2=\cos(x^3)\p_1+{\sin(x^3)\over\sin(x^1)}\p_2-\cot(x^1)\sin(x^3)\p_3$\\
$X_3=\p_3$
\end{tabular}

And also,

\begin{tabular}{cc}
$\omega^1=-\sin(x^3)dx^1+\sin(x^1)\cos(x^3)dx^2$; & $d\omega^1=\omega^2\wedge\omega^3$ \\
$\omega^2=\cos(x^3)dx^1+\sin(x^1)\sin(x^3)dx^2$; & $d\omega^2=\omega^3\wedge\omega^1$ \\
$\omega^3=\cos(x^1)dx^2+dx^3$; & $d\omega^3=\omega^1\wedge\omega^2$
\end{tabular}
\end{itemize}
For Types IV and VIII we give the structure constants only. For more explicit data on
these Types, see \cite{SR}
\begin{itemize}
\item {\bf{Type IV}}: $C^1_{13}=-C^1_{31}=1$, $C^1_{23}=-C^1_{32}=1$, $C^2_{23}=-C^2_{32}=1$
and rest $C^{i}_{j,k}=0$\\
\item{\bf{Type $\bf{VII_h \ (0<h^2<4)}$}}: $C^2_{13}=-C^2_{31}=1$, $C^1_{23}=-C^1_{32}=-1$, $C^2_{23}=-C^2_{32}=h$
and rest $C^{i}_{j,k}=0$\\
\item{\bf{Type VIII}}: $C^1_{23}=-C^1_{32}=-1$, $C^2_{31}=-C^2_{13}=1$, $C^3_{12}=-C^3_{21}=1$
and rest $C^{i}_{j,k}=0$\\
\end{itemize}

\section{Gauge Field Equation of Motion}\label{AppB}

In this appendix we will consider a system with action eq.(\ref{action}) and   metric   of form
\begin{equation}
\label{metappb}
 ds^2 = dr^2 - e^{2 \beta_t r} dt^2  + \eta_{ij}(r) \omega^i \omega^j
\end{equation}
and   derive the   equation
of motion for the gauge field. In eq.(\ref{metappb})  $\omega^i, i=1,2,3,$ are the  three invariant one-forms 
along the spatial directions in which the brane extends. 

To preserve the generalised translation symmetries  along the spatial directions 
 the gauge potential must take the form 
\begin{equation}
 A = \sum_if_{i}(r) \omega^i + A_t(r) dt.
\end{equation}

Eventually, we will take $\eta_{ij}$ to be diagonal
\be
\label{etaappb}
\eta_{ij}=( \lambda_1^2 e^{2 \beta_1 r}, \lambda_2^2 e^{2 \beta_2 r}, \lambda_3^2 e^{2\beta_3 r}) 
\ee
and the functions appearing in the gauge field to be  of the form
\be
\label{fungf}
f_i(r)=\tilde{A}_i e^{\beta_i r}, \ \ A_t(r)=A_t e^{\beta_t r}.
\ee
where $\lambda_i, \beta_i, \beta_t, \tilde A_i, A_t$ are all constants independent of all coordinates.
For now though, we keep them to be general and proceed. 

The gauge field equation of motion is 
\begin{equation}\label{gaugeeom}
 d *_5 F = - {1 \over 2} m^2 *_5 A.
\end{equation}

Only two cases are relevant for the  discussion above. Either the gauge field has 
  components only  along the spatial directions or only along time.
We discuss them in turn below.

\subsection{Gauge Field With Components Along Spatial Directions}

Using $d \omega^i ={1 \over 2} C^i_{\ jk} \omega^j \wedge
\omega^k $, we get the
 field strength to be\footnote{Note in our convention 
$F = {1\over 2} F_{\mu\nu} dx^\mu \wedge dx^\nu$.}
\begin{equation}
 F = d A = f'_i(r) dr \wedge \omega^i + {1 \over 2} f_i(r)
C^i_{\ jk} \ \omega^j \wedge \omega^k.
\end{equation}

With a choice of orientation so that in the basis $(\omega^1,\omega^2,\omega^3,dr, dt)$, $\epsilon_{123rt}>0,$
we have 
the  following Hodge dualities:
\begin{eqnarray}
 *_5 \omega^i &=&  {e^{\beta_t r} \over 2} \sqrt{\eta} \ \eta^{ij}
\epsilon_{jkl} \ \omega^k
\wedge \omega^l \wedge dr \wedge dt \\
*_5 (\omega^j \wedge \omega^k)  &=&  {e^{\beta_t r} \over  \sqrt{\eta} }
\eta_{il} \ 
\epsilon^{jkl} \ \omega^i \wedge dr \wedge dt\\
*_5 (dr\wedge \omega^i) &=& - {\sqrt{\eta}e^{\beta_t r} \over 2} \eta^{ij}
\epsilon_{jkl} \
\omega^k \wedge \omega^l \wedge dt.
\end{eqnarray}
Here we are using notation such  that $\epsilon^{ijk}=\epsilon_{ijk}=1$. 

Thus the R.H.S of the gauge field equation, eq(\ref{gaugeeom}), becomes,
\begin{equation}\label{stara}
-{1 \over 2 } m^2 *_5 A = - {1 \over 4 }m^2 f_i (r)
\eta^{ij} \sqrt{\eta}e^{\beta_t r} \epsilon_{jnp}  \ \omega^n \wedge
\omega^p \wedge dr \wedge dt.
\end{equation}
The L.H.S of the gauge field equation, eq(\ref{gaugeeom}), becomes,
\begin{eqnarray}\label{dstarf}\nonumber
 d*_5 F &=& d \left( -{\sqrt{\eta}e^{\beta_t r}\over 2}f'_i(r)  \eta^{ij}
\epsilon_{jkl} \  \omega^k \wedge \omega^l \wedge dt+ {e^{\beta_t r}
\over 2 \sqrt{\eta}} f_i(r)
C^i_{\ jk} \ \eta_{le} \ \epsilon^{ejk} \omega^l \wedge dr \wedge dt\right) \\
\nonumber
&=& \left[-{1 \over 2}\left(f'_i(r) e^{\beta_t r}\sqrt{\eta} \eta^{ij}\right)'
\epsilon_{jnp}+{1
\over 4\sqrt{\eta} } f_i(r)e^{\beta_t r} C^i_{\ jk} \eta_{le} C^l_{\ np}
\epsilon^{ejk} \right] \ \omega^n \wedge \omega^p \wedge dr \wedge dt\\
&&\ \ -{1 \over 2} f'_i(r)e^{\beta_t r} \sqrt{\eta} \eta^{ij} \epsilon_{jkl}
C^k_{\ qn} \
\omega^q \wedge \omega^n \wedge \omega^l \wedge dt.
\end{eqnarray}

Comparing eq.(\ref{stara}), eq.(\ref{dstarf}) we see that 
\begin{equation}
 f'_i (r) e^{\beta_t r} \sqrt{\eta} \eta^{ij}\epsilon_{jkl} C^k_{\
qn} \epsilon^{qnl} = 0.
\end{equation}
Defining 
\begin{eqnarray}
\label{defcab}
\epsilon^{ijl} C^k_{\ ij} &  \equiv & 2 C^{lk} \\
 \epsilon_{ijk} C^{ij} &  \equiv  &  2 a_k
\end{eqnarray}
  as in \cite{LL}, we get the condition that
\begin{equation}\label{adeqzero}
\text{either} \quad   a_k = 0\hspace{5mm} \mbox{ or } \hspace{5mm}f'_i(r)e^{\beta_t r}\sqrt{\eta}
\eta^{ij}(r)  a_j = 0.
\end{equation}

Here we consider the case when $a_k=0$, this includes Type I, II, III, VII$_0$, VIII, IX. In particular
it covers  all cases discussed in the main text   where  $A$ is oriented 
along the spatial directions. 
 
Now  comparing the $\omega \wedge \omega \wedge dr \wedge dt $ terms in eq.(\ref{stara}), eq.(\ref{dstarf}) and
multiplying by
$\epsilon^{mnp}$ on both sides  we get 
\begin{equation}
\label{gfeomappb}
 {1 \over 2} m^2 \sqrt{\eta} f_i(r)e^{\beta_t r} \eta^{im}
=(f'_i(r)e^{\beta_t r}
\sqrt{\eta} \eta^{im})' - {e^{\beta_t r} \over  \sqrt{\eta}} f_i(r)
C^{ji}C^{ml} \eta_{lj}.
\end{equation}

To proceed let us  consider  the metric  to be of form eq.(\ref{etaappb}). Also note that the gauge field is assumed 
to be of form as in eq.(\ref{fungf}) but with $A_t=0$. Also we take $C^{ji}$ to be diagonal of form
\be
\label{evcab}
C^{ji} = \delta^{ji} k^{j} \quad  \mbox{(with no sum over the index $j$),}
\ee
for some constants $k^j$ such that $2 a_k = \epsilon_{ijk} C^{ij} = 0$. 
This is equivalent to  
\be
\label{evcijk}
C^i_{\ jk} = \epsilon_{ijk} k^{i} \quad  \mbox{(with no sum over the index $i$).}
\ee
Then eq.(\ref{gfeomappb}) says that for every value of the  index $i=1,2,3,$ such that $\tilde{A}_i$ is non-vanishing, following two conditions must be met:
\be
\label{condoneb}
\sum_j \beta_j =  2 \beta_i 
\ee
and 
\be
\label{condtwob}
\left[ {1 \over 2}  m^2 -\beta_i (- \beta_i + \beta_t +
\sum_j \beta_j) \right]
 =   -  {\lambda_i^4 \over \ti \lambda^2} (k^i)^2,
\ee
where $\ti \lambda = \lambda_1 \lambda_2 \lambda_3$.

\subsection{Gauge Field With Components Only Along Time}
Next take the case where the gauge field has only a component along the time direction, 
\begin{equation}
 A = A_t(r) dt.
\end{equation}
The field strength becomes
\begin{equation}
 F=dA=A'_t(r) dr \wedge dt.
\end{equation}
Using the  Hodge star relations, 
\begin{eqnarray}
*_5 (dr\wedge dt) &=& - {\sqrt{\eta}e^{-\beta_t r} \over 6} \epsilon_{ijk} \
\omega^i \wedge \omega^j \wedge \omega^k \\
*_5 (dt) &=& {\sqrt{\eta}e^{-\beta_t r} \over 6} \epsilon_{ijk} \  dr \wedge
\omega^i \wedge \omega^j\wedge \omega^k,
\end{eqnarray}
 we get
\begin{eqnarray}
 d*_5 F&=&-{\epsilon_{ijk} \over 6} \left(\sqrt{\eta}e^{-\beta_t r} A'_t(r)\right)' \  dr \wedge
\omega^i \wedge \omega^j\wedge \omega^k\\
*_5 A&=&A_t(r){\sqrt{\eta}e^{-\beta_t r} \over 6} \epsilon_{ijk} \  dr \wedge
\omega^i \wedge \omega^j \wedge \omega^k.
\end{eqnarray}
Thus the gauge field equation of motion   eq.(\ref{gaugeeom}) becomes
\begin{equation}
 \left(\sqrt{\eta}e^{-\beta_t r} A'_t(r)\right)'={m^2\over2}A_t(r)\sqrt{\eta}e^{-\beta_t r}.
\end{equation}
With a metric eq.(\ref{etaappb}) and  gauge field
\be
\label{gfappb3}
A_t(r)=A_t e^{\beta_t r} ,
\ee
we get
\begin{equation}
 2 \beta_t\sum_i \beta_i=m^2,
\end{equation}
where $\ti \lambda=\lambda_1 \lambda_2 \lambda_3$.

\section{Lifshitz Solutions}\label{AppC}
In this appendix we examine Lifshitz solutions which are known to arise in the system, eq.(\ref{action}). 
Let us consider an ansatz,
\begin{equation}
 ds^2 = dr^2-e^{2 \beta_t r} dt^2+e^{2 \beta_i r}(dx^i)^2
\end{equation}
where $i=1,2,3$. Let us first turn on the gauge field along time direction,
\begin{equation}
 A=\sqrt{A_t} e^{\beta_t r} dt.
\end{equation}
The Maxwell equation gives
\begin{equation}
 m^2=2 \beta_t (\beta_1+\beta_2+\beta_3)
\end{equation}
and the trace reversed Einstein equations give
\begin{eqnarray}
A_t \beta_t^2 -3 (\beta_t^2+\beta_1^2+\beta_2^2+\beta_3^2)+\Lambda &=& 0\\
A_t (3 m^2 +4 \beta_t^2)-12 \beta_t (\beta_t+\beta_1+\beta_2+\beta_3)+ 4 \Lambda &=& 0\\
A_t\beta_t^2+6 \beta_1 (\beta_t+\beta_1+\beta_2+\beta_3) -2 \Lambda &=& 0 \\
A_t\beta_t^2+6 \beta_2 (\beta_t+\beta_1+\beta_2+\beta_3) -2 \Lambda &=& 0 \\
A_t\beta_t^2+6 \beta_3 (\beta_t+\beta_1+\beta_2+\beta_3) -2 \Lambda &=& 0.
\end{eqnarray}
The last three equations show $\beta_i=\beta$ for all $i$. Using this in the Maxwell equation we get
\begin{equation}
 m^2 = 6 \beta \beta_t.
\end{equation}
Then solving the other equation of motion gives the solution as
\begin{eqnarray}
\beta_i &=& \beta ~\forall~i \\
m^2 &=& 6 \beta \beta_t\\
\Lambda &=& \beta_t^2+2 \beta \beta_t + 9 \beta^2\\
A_t &=& 2 (1-\f{\beta}{\beta_t}).
\end{eqnarray}
If $\beta_t>0, \beta_i>0$,then $m^2$ is always positive. For the gauge field to be real, $\beta_t>\beta$.\\

Now, with the same metric ansatz, we choose the gauge field to be oriented along any one of $x^i$ directions.
Without loss of generality, let us choose it to be oriented along $x^1$:
\begin{equation}
 A=\sqrt{A_1} e^{\beta_1 r} dx^1.
\end{equation}
The equation of motion can be obtained from the previous case by $\beta_t \leftrightarrow \beta_1$ and $A_t \to -A_1$.
Then solving the equation of motion gives
\begin{eqnarray}
\label{lastlif}
\beta_t=\beta_i &=& \beta ~\forall~i\ne1 \\
m^2 &=& 6 \beta \beta_1\\
\Lambda &=& \beta_1^2+2 \beta \beta_1 + 9 \beta^2\\
A_1 &=& 2 (\f{\beta}{\beta_1}-1).
\end{eqnarray}
Again, if $\beta_t>0, \beta_i>0$, then $m^2$ is always positive. For the gauge field to be real $\beta>\beta_1$.

\section{Extremal RN solution}\label{AppD}

Starting  with the  action eq.(\ref{action}), and setting  $m^2=0$,  one gets the well known Reissner Nordstrom
black brane solution.

It has the metric
\begin{equation}
 ds^2=-a(\tir)dt^2 + {1\over a(\tir)} d\tir^2 +b(\tir)[dx^2+dy^2+dz^2]
\end{equation}
with
\begin{equation}
 a(\tir)= {Q^2 \over 12 \tir^4} + {\tir^2 \Lambda \over 12} - {M \over \tir^2}
\hspace{10mm} b(\tir)=\tir^2
\end{equation}
and the gauge field,
\begin{equation}
 A = -\left({Q  \over2 \tir^2}-{Q  \over2 \tir_h^2}\right)dt.
\end{equation}
Here $Q$ is charge and $M$ is the mass of the black brane. 

In the extremal limit we get $a'(\tir_h)=0$ and also $a(\tir_h)=0$.
This allows us to solve for $Q$ and $M$ in terms of $\tir_h$,
\begin{eqnarray}
 M &=& \frac{\tir_h^4 \Lambda}{4}\\
Q^2 &=& 2 \tir_h^6 \Lambda,
\end{eqnarray}
giving eq.(\ref{metern}), eq.(\ref{gfern2}).

Now consider a new radial coordinate, $\tilde{r}$, given by eq.(\ref{defrtilde}).. 
The relation between the coordinates near $\tir=\tir_h$ is given by
\begin{equation}
 \sqrt{\Lambda} r= \log\left(\frac{\tir-\tir_h}{\tir_h}\right)+\frac{7}{6}
\frac{\tir-\tir_h}{\tir_h}+\cdots.
\end{equation}
This can be inverted  (as $\tir \to \tir_h$, $r \to -\infty$) to give
\begin{equation}
{\ti r - \ti r_h  \over \ti r_h } = e^{\sqrt{\Lambda} r} \left[ 1 - {7 \over 6}
e^{\sqrt{\Lambda} r} + \cdots \right].
\end{equation}
In the new coordinates the metric near $r=-\infty$ becomes
\begin{eqnarray}
 ds^2 &=&  dr^2
- \tir_h^2 \Lambda e^{2 \sqrt{\Lambda} r} \left(1-\frac{14}{3}e^{\sqrt{\Lambda} r}
+\cdots\right)dt^2 \nonumber \\
&+& \tir_h^2 \left(1+2 e^{\sqrt{\Lambda} r} +\cdots\right) (dx^2+dy^2+dz^2)
\end{eqnarray}
and the gauge field becomes
\begin{equation}
 A= \tir_h \sqrt{\Lambda} \sqrt{2}e^{ \sqrt{\Lambda} r}\left(1-\frac{8}{3}
e^{\sqrt{\Lambda} r} +\cdots\right) dt.
\end{equation}
If we rescale the coordinates as $t\to \frac{t}{\tir_h \sqrt{\Lambda}}$ and
$\{x,y,z\} \to \frac{1}{\tir_h} \{x,y,z\}$,
then the solution up to first order in deviation near $r=-\infty$ can be written
as, eq.(\ref{metint}), eq.(\ref{gint}).

%%%%%%%%%%%%%%%%%%%%

\end{document}